\documentclass[preprint,12pt]{elsarticle} 
\usepackage[utf8]{inputenc}
\usepackage{graphicx}
\usepackage{varwidth}
\usepackage{amsmath}
\usepackage{txfonts}
\usepackage{cases}
\usepackage{subfig}
\usepackage{float}
\usepackage{mathtools}
\usepackage{fullpage}
\usepackage{comment}
\usepackage{algorithmicx}
\usepackage{algorithm}
\usepackage[noend]{algpseudocode}
\usepackage{siunitx}
\usepackage{xcolor}
\usepackage{setspace}
\usepackage{doi}
\usepackage{url}
\usepackage{epstopdf}
\biboptions{compress}
\usepackage[symbol]{footmisc}
\def\correspondingauthor{\footnote{Corresponding author. Benjamin Werner \href{mailto:benjamin.werner@cvut.cz}{benjamin.werner@cvut.cz}}}

\begin{document}

\begin{frontmatter}

\title{Extended Quasicontinuum Methodology for Highly Heterogeneous Discrete Systems}

\author[1]{Benjamin Werner \correspondingauthor{}}
\author[2]{Ond\v{r}ej Roko\v{s}}
\author[1]{Jan Zeman}

\address[1]{Faculty of Civil Engineering, Czech Technical University in Prague, Th\'{a}kurova 7, 166 29 Prague 6, Czech Republic}

\address[2]{Mechanics of Materials, Department of Mechanical Engineering, Eindhoven University of Technology, P.O. Box 513, 5600 MB Eindhoven, The Netherlands}

\begin{abstract}
Lattice networks are indispensable to study heterogeneous materials such as concrete or rock as well as textiles and woven fabrics. Due to the discrete character of lattices, they quickly become computationally intensive. The QuasiContinuum (QC) Method resolves this challenge by interpolating the displacement of the underlying lattice with a coarser finite element mesh and sampling strategies to accelerate the assembly of the resulting system of governing equations. In lattices with complex heterogeneous microstructures with a high number of randomly shaped inclusions the QC leads to an almost fully-resolved system due to the many interfaces. In the present study the QC Method is expanded with enrichment strategies from the eXtended Finite Element Method (XFEM) to resolve material interfaces using nonconforming meshes. The goal of this contribution is to bridge this gap and improve the computational efficiency of the method. To this end, four different enrichment strategies are compared in terms of their accuracy and convergence behavior. These include the Heaviside, absolute value, modified absolute value and the corrected XFEM enrichment. It is shown that the Heaviside enrichment is the most accurate and straightforward to implement. A first-order interaction based summation rule is applied and adapted for the extended QC for elements intersected by a material interface to complement the Heaviside enrichment. The developed methodology is demonstrated by three numerical examples in comparison with the standard QC and the full solution. The extended QC is also able to predict the results with \SI{5}{\%} error compared to the full solution, while employing almost one order of magnitude fewer degrees of freedom than the standard QC and even more compared to the fully-resolved system.
\end{abstract}
\end{frontmatter}

\section{Introduction}

Numerical models of discrete lattice networks are beneficial to study the failure behavior of various heterogeneous materials. These numerical models are simple to assemble and represent the underlying microstructure directly. They are highly suitable as numerical models for cohesive-frictional materials such as concrete and rock \cite{Nikolic2018753,NIKOLIC2015209}. The lattice nodes represent center points of grains, connected by spring or beam elements and are arranged as regular or irregular lattices \cite{Nikolic2018753,BOLANDER2021108030}. A great advantage of these models is their numerical robustness in predicting crack growth and crack coalescence in quasi-brittle materials \cite{Elias2015,Chang20204822} due to their local representation of the microstructure. Continuum models, in contrast, are based on homogenized material behavior and lead to an element size-dependency of strength and fracture toughness predictions, stress-locking problems, and difficulties to capture multiple growing cracks \cite{Lilliu2007, BOLANDER2021108030}. Over the past decades the approach of using lattice networks to study numerically quasi-brittle materials such as concrete has been continuously improved by inelastic damage behavior of the spring elements connecting the particles as well as friction between them \cite{Grassl2010,Cusatis2006}. Discrete lattice networks have also been applied to study fracture behavior of cortical bovine bone \cite{Mayya2016} or textiles and woven fabrics \cite{Wilbrink2013,Beex2013}. The elements of the lattice are representing yarns and fibers of textiles or fabrics at the meso-scale level and thus directly allow to incorporate bond failure in a straightforward manner. The fibrous material of paper has been represented by lattice networks in numerical studies as well, and shown to reproduce the fracture processes accurately \cite{Liu2010}.

\begin{figure}[ht]
	\centering
	\includegraphics[width=0.5\textwidth]{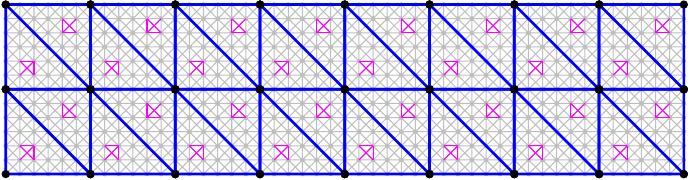}
    \caption{ \small{A schematic illustration of a QC model with an underlying x-braced lattice in grey, a finite element mesh in blue, repatoms in black and sampling interactions in magenta.}}
    \label{fig:QC_IntSum}
\end{figure}

Discrete lattices provide a simple way to investigate heterogeneous materials, but they are usually costly to simulate. The QuasiContinuum (QC) method is a numerical multi-scale approach for lattice networks addressing that problem. This numerical method was first introduced by Tadmor et al. \cite{Tadmor1996,Tadmor1996a} and proposed for crystal lattices on the atomistic level to investigate dislocations and nanoindentation. The methodology relies on two main steps. (i) Instead of considering all degrees of freedom (DOF) of the full atomistic arrangement the system of equations in QC is reduced by \emph{interpolation}. Here, a subset of so-called representative atoms (repatoms) is selected for a triangulation or discretization (Figure \ref{fig:QC_IntSum}) with a linear interpolation to approximate the displacements of the full atomistic arrangement, in the spirit of the Finite Element Method (FEM). (ii) To avoid visiting every lattice site to determine the total potential energy of the system, and to assemble the resulting system of governing equations, \emph{summation} rules are used to further reduce the computational cost of the method. A certain set of sampling atoms or interactions is selected (Figure \ref{fig:QC_IntSum}) with corresponding weight factors to approximate the total potential energy of the entire system. The weight factors are thereby indicating the number of sites the sampling atom is representing in a triangle, in analogy to the weights used in Gauss integration values. With these two integral parts the QC method allows studying solids at the atomistic level or microscale at locations with highly non-uniform deformations, while seamlessly transitioning to the continuum or macroscopic level at places with small deformation gradients with limited computational efforts \cite{Miller2002,Curtin2003}.

The QC methodology has been applied to problems at different length scales, including simulations of nano cutting of a copper crystal by Yang et al. \cite{Yang2021}. Here, the QC has been extended with a material removal criterion to analyse the ultra-precision process more accurately and to determine the influence of various parameters on the cutting force. Tran et al. \cite{Tran2019} applied the method at the micro-scale to study the friction behaviour of textured and rough aluminum surfaces. A cylindrical indenter was moved along the surface and the friction coefficient was determined, among other parameters. 

Besides investigations of atomistic arrangements, QC has been applied to lattices at the macroscopic level. Beex et al. \cite{Beex2015} used the method to study fabric with elastic-plastic material behavior. Individual links of the lattice represent yarn segments of the fabric while the lattice nodes serve as yarn-to-yarn contact points. To include dissipative material behavior, the QC was reformulated in a virtual-power-based manner and extended to an interaction based summation rule in \cite{Beex2015}. Generalization to dissipative processes based on a variational formulation has been performed in \cite{Rokos2016}. The QC method holds potential for irregular lattice arrangements, which is of great advantage for use in concrete and rock and concrete analysis \cite{Mikes2017}. 

Further extensions involve beam lattice structures using the corotational beam element formulation allowing for geometrical nonlinearities. Phlipot and Kochmann \cite{Phlipot2018} presented two- and three-dimensional indentation tests and a mode I sample with a pre-existing crack. They used multiple lattice unit cell geometries including triangular, hexagonal and Body Centered Cubic (BCC) arrangements. Chen et al. \cite{Chen2021} studied elastic three-dimensional lattice structures with BCC and Kelvin unit cells and presented multiple numerical examples including contact definition between indentor and lattice structure. A more accurate refinement indicator for automatic adaptive meshing has also been developed. In an earlier study, Chen et al. \cite{Chen2020} extended the approach with corotational beam elements to plastic material behavior. Multiple beam elements with plastic hinges have been combined to represent one strut of the lattice to study indentation in a three-dimensional BCC arrangement. 

The interpolation step, when used with meshes, decrease the number of unknowns substantially, and is one of the integral parts of the QC method. Automatic adaptive refinement of the discretization is a further key to reduce the computational cost. During a numerical analysis the triangulation is refined at locations of high interest such as the contact area in indentation scenarios \cite{Knap2001,Chen2021}, the surrounding of crack tips \cite{Ghareeb2020,Rokos2017} or along inhomogeneities like grain boundaries \cite{Shenoy1999} or vacancies \cite{Memarnahavandi2015}. %

In the XFEM such discontinuities can be incorporate using nonconforming meshes \cite{Moes2003,Fries2010,Cheng2010,Li2018,Zhu2012186,WU201577,GUPTA201323} and are considered by additional DOFs and enriched interpolation functions. Roko{\v{s}} et al. \cite{Rokos2017x} used enrichments from XFEM to model crack propagation in discrete lattice networks and expanded the idea of computational efficiency through automatic mesh adaptivity in the QC method to the coarsening of the discretization. By including enriched interpolation functions the mesh is coarsened in the wake of the crack while the area around the tip of the propagating crack in a dissipative lattice networks is fully resolved. This leads to a significantly reduced number of DOFs while keeping high accuracy. To incorporate grain boundaries, vacancies and voids, material interfaces or other heterogeneities in numerical models they have to be fully resolved by the discretization in the current state of the QC method. Including enriched interpolation functions for weak discontinuities from the XFEM would allow to use a nonconforming triangulation. Consequently, the mesh does not have to be refined at the interface, further decreasing the number of unknowns in the system of equations significantly. 

In the energy-based QC method, evolution of the entire system is governed by the total potential energy and its accuracy depends on the applied summation rule. Therefore, a significant effort has been spent on defining suitable summation rules which introduce a minimal error and are at the same time computationally efficient. In the first studies developing the QC method the atom closest to the quadrature point of the triangle was selected as the sampling atom \cite{Tadmor1996,Tadmor1996a}. This resulted in a computationally efficient summation rule for large triangles but the selection procedure does not transition seamlessly to fully resolved regions experiencing high deformations. As a result, non-physical forces appear at the interface between the fully and non-fully resolved domains and introduce errors in the analysis. The cluster based summation rule of Knap and Ortiz \cite{Knap2001} resolves this problem by considering multiple sampling nodes located in a circle around the repatoms. Eidel and Stukowski \cite{Eidel2009} investigated the cluster summation rule in more detail and proposed strategies to further reduce the error in the approximated potential energy. Yang et al. \cite{Yang2013} combined cluster and quadrature point summation rule. A similar approach is the central summation rule by Beex et al. \cite{Beex2014} where the cluster is reduced to the repatom itself and the remaining atoms are accounted for by the center atom. In an earlier study Beex et al. \cite{Beex2011} presented an exact summation rule whereby every atom with a neighbour in an adjacent triangle is accounted for individually for sampling and the center atom is representing the remaining atoms inside the triangle. This leads to small bands of sampling atoms along the triangle edges, thus resulting in higher accuracy but requires more computational effort. Amelang et al. \cite{Amelang2015} as well as Amelang and Kochmann \cite{Amelang2015a} studied and systematically summarized existing summation rules. These studies provide optimal summation rules for non-uniform meshes for one-, two- and three-dimensional lattices. The selection procedure of sampling atoms is similar to the studies of Beex et al. \cite{Beex2014} and Yang et al. \cite{Yang2013} as a combination of a cluster at the repatoms combined with a sampling atom at the quadrature point. For higher order interpolation functions multiple quadrature points inside the triangle and additional sampling atoms on the edges are applied. Instead of using sampling atoms to approximate the potential energy of the system, Beex et al. \cite{Beex2015} used the interactions between them. In an X-braced lattice one link for each of the four directions is selected to represent all of them in a triangle. In addition, links crossing triangle edges are accounted discretely in the summation rule which leads to an exact assembly of the potential energy.

The main goal of the present study is incorporating weak discontinuities into the QC method by using enriched interpolation functions motivated by the XFEM to account for interfaces in heterogeneous discrete systems, to further reduced number of DOFs and speed up calculations. To quantify the error of the approach, the outcome is compared to the standard QC with fully-resolved discontinuity boundaries as well as the underlying full direct numerical solution. Three different numerical examples are investigated. The first two include a single stiff circular inclusion and a single stiff fiber embedded in a soft matrix and studied under tensile loading. The third example represents a cross-section of a concrete sample with multiple aggregates and fibers which are surrounded in a more compliant cement matrix, studied under tensile loading using Periodic Boundary Conditions (PBC). For the extended as well as the standard QC the interaction based first-order summation rule is applied and compared to the full summation. The main outcome of this study is the convergence behavior of the standard QC and the extended QC with four different enrichment types (Heaviside, absolute value, modified absolute value, and corrected XFEM) using the full summation. The convergence of both QC approaches using the interaction based first-order summation rule is demonstrated. Moreover, the accuracy and the numerical efficiency of the extended and the standard QC is compared for all three numerical examples.

\section{Methods}
\subsection{Full-Lattice Formulation}
\begin{figure}[ht]
	\centering
	\begin{minipage}{0.25\linewidth}
		\begin{center}
			\includegraphics[scale=1]{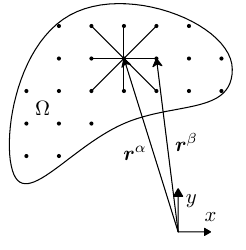}
		\end{center}
		\vspace{-1em}
		\footnotesize{global kinematic variable: \\
			$\mathbf{r}=[\mathbf{r}^1,\dots,\mathbf{r}^{n_\mathrm{ato}}]^\mathsf{T}$}
	\end{minipage}
	\hspace{0.1\linewidth}
	\begin{minipage}{0.4\linewidth}	
		\begin{center}
			\includegraphics[scale=1]{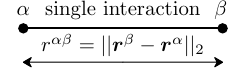}
		\end{center}
		\vspace{-0.9em}
		\footnotesize{elastic strain: $\varepsilon^{\alpha\beta}(r^{\alpha\beta}) = (r^{\alpha\beta}-r_0^{\alpha\beta})/r_0^{\alpha\beta}$}\\
	\end{minipage}
	\caption{Schematic illustration of the kinematic variables of an X-braced lattice.}
	\label{fig:Schematic_Kinematic}
\end{figure}

The lattice network considered is an X-braced arrangement in two dimensions where one atom is connected by interactions to eight direct neighbours inside the domain $\Omega$ (Figure \ref{fig:Schematic_Kinematic}). The location of each atom is described by a position vector and the position vectors of all atoms are stored in a single column matrix $\mathbf{r} = [\mathbf{r}^1,\ldots,\mathbf{r}^{n_\mathrm{ato}}]^T$ for the entire system with $n_\mathrm{ato}$ atoms. An equilibrium configuration $\mathbf{r}$ of the discrete system is determined by the minimization of the total potential energy $\Pi^k$ at a time instant $t_k$ as 
\begin{equation}
    \mathbf{r}(t_{k}) = \underset{\widehat{\mathbf{r}} \in \mathcal{R}^{2\, n_{\mathrm{ato}}}}{\mathrm{arg\,min\,}} \Pi^k(\widehat{\mathbf{r}}), \quad k=1,\ldots,n_{\mathrm{T}},
\end{equation}
where $\widehat{\mathbf{r}}$ denotes an arbitrary admissible position vector, and with the initial condition $\mathbf{r}(0) = \mathbf{r}_0$ and $0 = t_0 < t_1 < \ldots < t_{n_\mathrm{T}} = T$ being discretization of a considered  time interval $[0,T]$. The potential energy of the lattice
\begin{equation}
    \Pi^k(\widehat{\mathbf{r}}) = \sum \limits_{\alpha \beta = 1}^{n_{\mathrm{int}}} \pi^k_{\alpha\beta} (\widehat{\mathbf{r}})
\end{equation}
is the sum of all elastic energies $\pi^k_{\alpha \beta}$ of all interactions at time $t_k$. The entire system contains $n_{\mathrm{int}}$ interactions stored in an index set $N_{\mathrm{int}}$. Each interaction $\alpha\beta \in N_{\mathrm{int}}$ is connecting the lattice sites $\alpha$ and $\beta$ with the length 
\begin{equation}
    r^{\alpha\beta} = ||\mathbf{r}^{\beta} - \mathbf{r}^{\alpha}||_2,
\end{equation} 
determined by the Euclidean norm of the difference between the two position vectors $\mathbf{r}^{\alpha}$ and $\mathbf{r}^{\beta}$. The elastic energy of the link between lattice sites $\alpha$ and $\beta$ is defined as
\begin{equation}
    \pi^k_{\alpha\beta} (\widehat{\mathbf{r}}) = \dfrac{E^{\alpha\beta}A^{\alpha\beta}}{2r_0^{\alpha\beta}}(\widehat{r}^{\alpha\beta}-r_0^{\alpha\beta})^2
\end{equation}
determined from the Young's modulus $E^{\alpha\beta}$, the link cross-section area $A^{\alpha\beta}$ as well as the strain of the link, which is the difference between the current $\widehat{r}^{\alpha\beta}$ and the initial length $r^{\alpha\beta}_0$ of the interaction, divided by the initial length.

After determining the potential energy of the whole lattice arrangement, the system of equations can be solved iteratively using Newton-Raphson method. The potential energy function is thereby approximated by the Taylor expansion, resulting in the following iterative systems of equations 
\begin{equation}
    \mathbf{K}^i(\widehat{\mathbf{r}}^{i+1}-\widehat{\mathbf{r}}^i) + \mathbf{f}^i = \mathbf{0}
    \label{eq:TaylorExpansion}
\end{equation}
where the interval force vector is defined as the gradient of $\Pi^k$,
\begin{equation}
    \mathbf{f}^i = \mathbf{f}(\widehat{\mathbf{r}}^i) = \left. \dfrac{\partial \Pi^k(\widehat{\mathbf{r}})}{\partial \widehat{\mathbf{r}}} \right|_{\widehat{\mathbf{r}}=\widehat{\mathbf{r}}^i}
\end{equation}
and the tangent stiffness matrix as the Hessian of $\Pi^k$,
\begin{equation}
    \mathbf{K}^i = \mathbf{K}(\widehat{\mathbf{r}}^i) = \left. \dfrac{\partial^2 \Pi^k(\widehat{\mathbf{r}})}{\partial \widehat{\mathbf{r}}\partial\widehat{\mathbf{r}}}   \right|_{\widehat{\mathbf{r}}=\widehat{\mathbf{r}}^i},
\end{equation}
both determined at a time instance $t_k$. Solving Eq. (\ref{eq:TaylorExpansion}) provides the location of the current increment of all position vectors. The iteration continuous until the combination of the normalized position vector and the normalized force is smaller than a predefined tolerance $\mathrm{tol}$, i.e.,
\begin{equation}
\dfrac{\lvert\lvert (\widehat{\mathbf{r}}^{i+1}-\widehat{\mathbf{r}}^i) \rvert\rvert}{\lvert\lvert \mathbf{r}_{\mathrm{free}}\rvert\rvert}  + \dfrac{\lvert\lvert \mathbf{f}^i \rvert\rvert}{\lvert\lvert \mathbf{f}_{\mathrm{DBC}}\rvert\rvert} < \mathrm{tol}.
\label{eq:tolerance}
\end{equation}
In Eq. (\ref{eq:tolerance}) the entire column matrix is split according to free and prediscribed DOFs $\mathbf{r} = [\mathbf{r}_{\mathrm{free}}, \mathbf{r}_{\mathrm{DBC}}]$ with $\mathbf{r}_{\mathrm{DBC}}$ being the position vectors of the atoms with fixed displacements and $\mathbf{r}_{\mathrm{free}}$ atoms left to move freely. The global column matrix of the internal forces $\mathbf{f} = [\mathbf{f}_{\mathrm{free}}, \mathbf{f}_{\mathrm{DBC}}]$ is split equivalently and the reaction forces of the atoms with prescribe displacements, $\mathbf{f}_{\mathrm{DBC}}$, are used in the convergence criterion for normalization.

\subsection{QC Reduction by Interpolation}
\label{sec:Interpolation}

To reduce the number of unknowns in Eq. (\ref{eq:TaylorExpansion}), atoms displacements are interpolated. The interpolation in the present study contains two parts, the standard FEM interpolation by using a triangular mesh, and an enriched interpolation motivated by the XFEM. Only the first part is applied in standard QC Method, whereas a combination of both results in the extended QC. The standard QC interpolation for the position vector of atom $\alpha$,
\begin{equation}
    \widehat{\mathbf{r}}^{\alpha}_{\mathrm{qc}} = \sum_{\beta \in N_{\mathrm{rep}}} \phi_{\beta}(\mathbf{r}_0^{\alpha})\widehat{\mathbf{r}}_{\mathrm{rep}}^{\beta}, \quad \alpha \in N_{\mathrm{ato}},
    \label{eq:interpoaltion_standard}
\end{equation}
is determined by the FE shape function $\phi_{\beta}(\mathbf{r}_0^{\alpha})$ and the location of the repatoms $\widehat{\mathbf{r}}_{\mathrm{rep}}^{\beta}$ with $N_{\mathrm{ato}}$ and $N_{\mathrm{rep}}$ being index sets containing all lattice atoms and repatoms. 

\begin{figure}
    \centering
    \subfloat[Heaviside]{\includegraphics[width=0.5\textwidth]{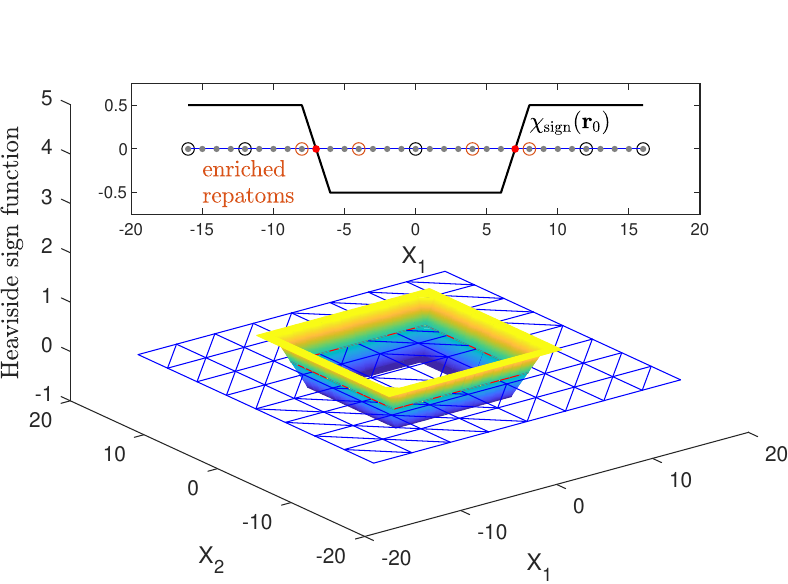}}
    \subfloat[absolute value]{\includegraphics[width=0.5\textwidth]{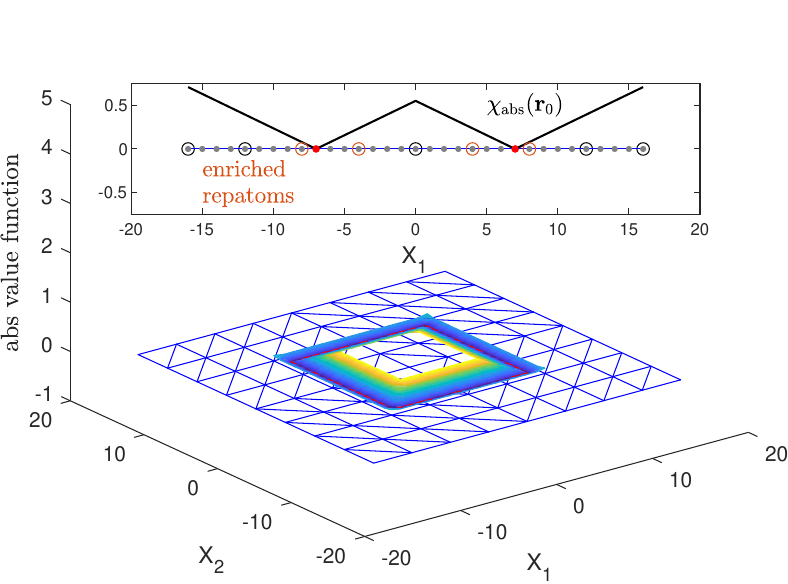}}\\
    \subfloat[modified absolute value]{\includegraphics[width=0.5\textwidth]{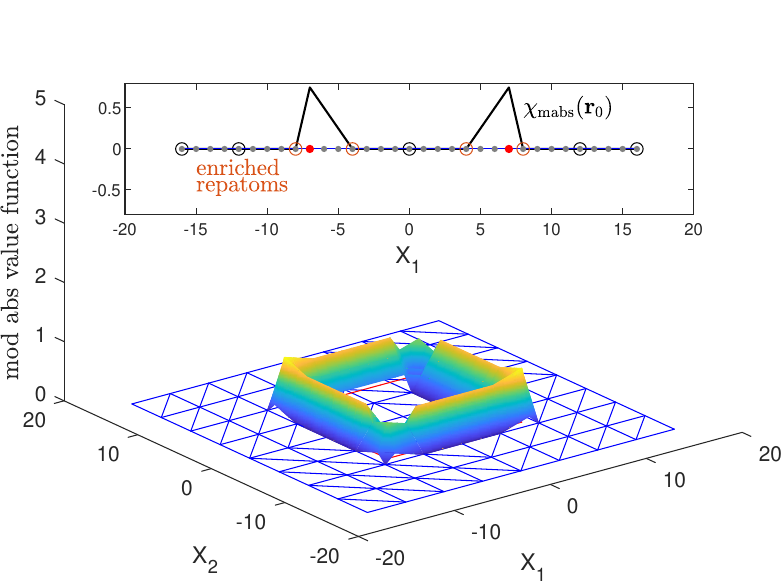}}
    \subfloat[corrected XFEM]{\includegraphics[width=0.5\textwidth]{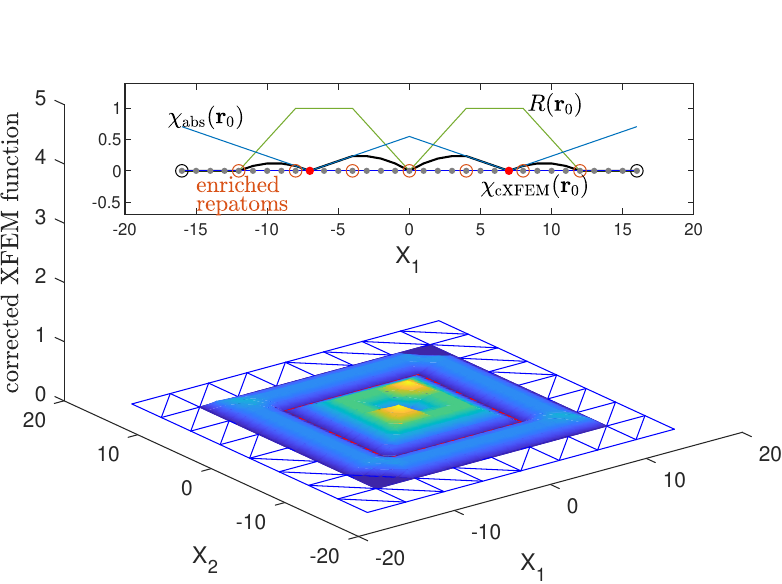}}
    \caption{The four enrichment functions (a) Heaviside, (b) absolute value, (c) modified absolute value, and (d) corrected XFEM, corresponding to a square inclusion and highlighted in red. In addition, cross-sections of individual enrichments are visualized as 2D curves along the $X_1$-axis and for $X_2=0$ in the insets. The insets include the repatoms as open circles, the atoms as grey dots as well as the atom on the interface indicated in red dots.}
    \label{fig:ExampleEnrichmentFunc}
\end{figure}

For the extended QC, four different enrichment functions are applied to account for weak discontinuities within a domain. This includes the Heaviside function \cite{Moes1999}, the absolute value enrichment \cite{Krongauz1998}, the modified absolute value enrichment \cite{Moes2003} as well as the corrected XFEM \cite{Fries2008}. They will be established and explained in the following. Before we introduce these four enrichment options, the general approximation is defined here. The position vector for each atom in the lattice,
\begin{equation}
    \widehat{\mathbf{r}}^{\alpha} = \widehat{\mathbf{r}}_{\mathrm{qc}}^{\alpha} +
    \sum_{j=1}^{n^{\star}} 
    \underbrace{
    \phi_{\beta_j}(\mathbf{r}_0^{\alpha})R(\mathbf{r}_0^{\alpha})[\chi_{\bullet}(\mathbf{r}_0^{\alpha}) - \chi_{\bullet}(\mathbf{r}_0^{\beta_j})]
    }_{\phi_j^{\star}(\mathbf{r}_0^{\alpha})}
    \widehat{\mathbf{g}}_{j}^{\star}, \quad \alpha \in N_{\mathrm{ato}}, \quad \beta_j \in N^{\star}
    \label{eq:interpoaltion_enrichment}
\end{equation}
is determined by the superposition of the standard interpolation $\widehat{\mathbf{r}}^{\alpha}_{\mathrm{qc}}$ from Eq. (\ref{eq:interpoaltion_standard}) and the enrichment. The enriched interpolation is calculated by the shape function $\phi_{\beta_j}(\mathbf{r}_0^{\alpha})$ with $\beta_j$ being one of the enriched repatoms $n^{\star}$ in the index set $N^{\star}$. The ramp function $R(\mathbf{r}_0^{\alpha})$ is only relevant for the corrected XFEM (explained later) and is shown in Figure \ref{fig:ExampleEnrichmentFunc}d; for the other three enrichment types it can be assumed equal to one for all atoms. The enriched degrees of freedom $\widehat{\mathbf{g}}_{j}^{\star}$ additionally account for the interfaces inside elements. The enrichment function $\chi_{\bullet}(\mathbf{r}_0^{\alpha})$ is one of the four options, explained in the following, and applied in its shifted form to ensure the Kronecker-delta property (the shift is guaranteed by the second term of the square bracket in Eq. (\ref{eq:interpoaltion_enrichment})). That leads to a vanishing of the enrichment functions in all repatoms except the ones belonging to an element with the interface and limits the additional DOFs to the minimum. In addition, the enrichment functions are zero along the boundaries, which greatly simplifies application of boundary conditions.

\begin{figure}
    \centering
    \subfloat[Triangulation]{\includegraphics[width=0.4\textwidth]{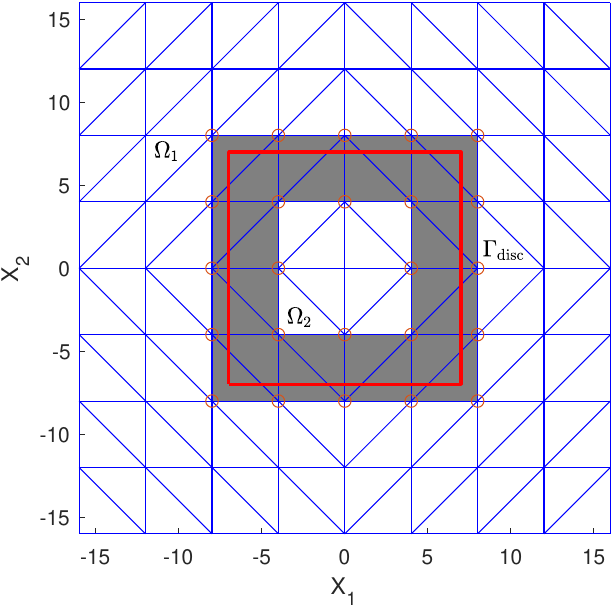}}
    \hfill
    \subfloat[Signed distance function]{\includegraphics[width=0.55\textwidth]{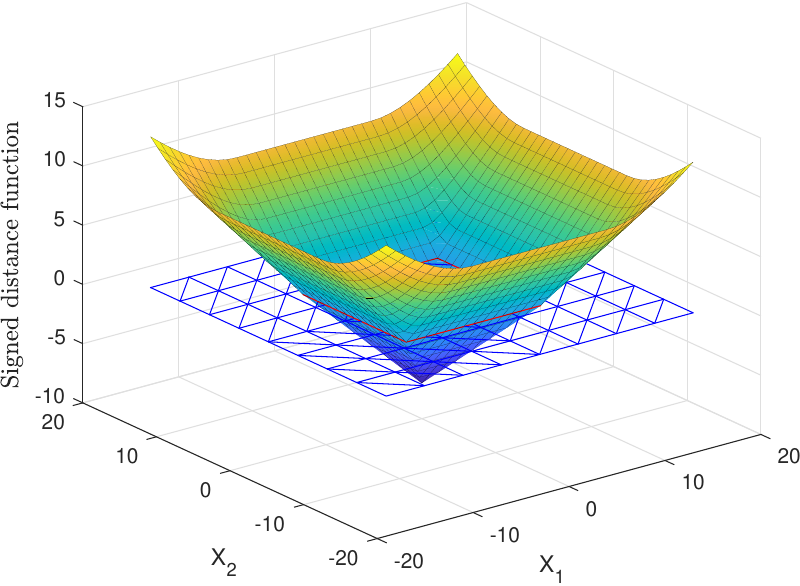}}
    \caption{(a) Example with a square inclusion and repatoms with the discontinuity $\Gamma_{\mathrm{disc}}$ separating the two domains $\Omega_{1}$ (compliant matrix) and $\Omega_{2}$ (stiff inclusion) together with the enriched elements in dark grey; (b) Signed distance function of the discontinuity}
    \label{fig:SignDistFunc}
\end{figure}

The enrichment functions are based on the signed distance function 
\begin{equation}
    \psi(\mathbf{r}_0^{\alpha}) = \lvert\lvert\mathbf{r}_0^{\alpha} - \mathbf{c}^{\alpha}\rvert\rvert \cdot \mathrm{sign}(\mathbf{n}_{c^{\alpha}}^{\mathrm{T}}(\mathbf{r}_0^{\alpha} - \mathbf{c}^{\alpha}))
    \label{eq:SignDistFunc}
\end{equation}
with the Euclidean norm of the difference between the location $\mathbf{r}_0^{\alpha}$ of the atom $\alpha$ in the initial configuration and the closest location $\mathbf{c}^{\alpha}$ to $\alpha$ on the discontinuity $\Gamma_{\mathrm{disc}}$ (red line in Figure \ref{fig:SignDistFunc}a) \cite{Fries2010}. The second term in Eq. (\ref{eq:SignDistFunc}) determines the sign for the current atom $\alpha$ depending on the normal vector $\mathbf{n}_{c^{\alpha}}$ in $\mathbf{c}^{\alpha}$. This ensures different signs on both sides of $\Gamma_{\mathrm{disc}}$ and is shown in Figure \ref{fig:SignDistFunc}b for a square inclusion as a cone like surface.

The Heaviside sign function
\begin{equation}
    \chi_{\mathrm{sign}}(\mathbf{r}_0) = 
    \begin{cases}
        -0.5, & \psi(\mathbf{r}_0) < 0, \\
        0, & \psi(\mathbf{r}_0) = 0,\\
        +0.5, & \psi(\mathbf{r}_0) > 0,
    \end{cases}
    \label{eq:enrichment_Heav}
\end{equation}
switches from $+0.5$ to $-0.5$ through the enriched elements with zero at the interface and is the first enrichment type. It is commonly used for strong discontinuities such as cracks but can be applied to weak discontinuities as well. In that case the Heaviside sign function has to be zero along the interface \cite{Fries2010}. In analyses of continua using XFEM this condition is enforced by using the Lagrange multipliers or Nitsche's method \cite{Fries2010,Lang20141357,KRAMER201370}. Due to the discrete character of the underlying lattice network, this requirement can be easily applied by setting $\chi_{\mathrm{sign}}$ zero along the interface in Eq. (\ref{eq:enrichment_Heav}). Due to its constant character, the Heaviside enrichment in its shifted form has the advantage to vanish outside elements crossed by an interface. This avoids so called blending elements, characterized by only a subset of the elements nodes or repatoms are enriched, and can lead to suboptimal convergence behavior \cite{Fries2010}. In Figure \ref{fig:ExampleEnrichmentFunc}a, the enrichment function is shown in its unshifted form as a surface along the discontinuity together with the triangulation of the square inclusion example from Figure~\ref{fig:SignDistFunc}. Moreover, a cross section of the Heaviside function along the $X_1$-axis for $X_2=0$ is shown as well as the enriched repatoms (indicated as orange circles).  

The absolute value enrichment  
\begin{equation}
    \chi_{\mathrm{abs}}(\mathbf{r}_0) = \mathrm{abs}(\psi(\mathbf{r}_0))
    \label{eq:enrichment_abs}
\end{equation}
describes the location of the discontinuity by using the absolute value of the signed distance function and is zero along the discontinuity (Figure \ref{fig:ExampleEnrichmentFunc}b). It is a common enrichment function in the context of XFEM to investigate continua to account for weak discontinuities. A disadvantage of the absolute value enrichment are blending elements, which can lead to suboptimal results.

The third type of enrichment is the modified absolute value function. The interface inside the enriched elements is described by a ridge of the interpolation function (Figure \ref{fig:ExampleEnrichmentFunc}c). The modified absolute value enrichment 
\begin{equation}
    \chi_{\mathrm{mabs}}(\mathbf{r}_0) = \sum_{j=1}^{n^{\star}}\phi_{\beta_j}(\mathbf{r}_0)\cdot|\psi(\mathbf{r}_0^{\beta_j})| - \left| \sum_{j=1}^{n^{\star}}\phi_{\beta_j}(\mathbf{r}_0)\cdot\psi(\mathbf{r}_0^{\beta_j}) \right|
    \label{eq:enrichment_mabs}
\end{equation}
is calculated by the standard FE shape function $\phi_{\beta_j}(\mathbf{r}_0)$ at atom $\beta$ and the signed distance function $\psi$ at one of the enriched repatoms $n^{\star}$. It was first introduced by Mo\"{e}s et al. \cite{Moes2003} and addresses the suboptimal results of the absolute value function. All repatoms outside the enriched elements are zero and therefore blending elements are avoided.

Finally, the fourth enrichment function is the corrected XFEM
\begin{equation}
    \chi_{\mathrm{cXFEM}}(\mathbf{r}_0) = \mathrm{abs}(\psi(\mathbf{r}_0)) \cdot R(\mathbf{r}_0),
    \label{eq:enrichment_cXFEM}
\end{equation}
which is calculated by the absolute value function multiplied with a ramp function $R(\mathbf{r}_0)$. The product of the two functions leads to the enrichment function visualized as a surface plot and its cross section in 2D along the $X_1$-axis and for $X_2=0$ in Figure \ref{fig:ExampleEnrichmentFunc}d. The ramp function takes the value of one for all enriched repatoms and is zero for the remaining ones. Identical to the Heaviside and absolute value enrichment the corrected XFEM is characterized by zero values along the interface. Furthermore, equally to the modified absolute value enrichment the corrected XFEM function is zero for the repatoms of all non-enriched elements and blending elements do not appear.

\begin{figure}
    \centering
    \subfloat[]{\includegraphics[width=0.4\textwidth]{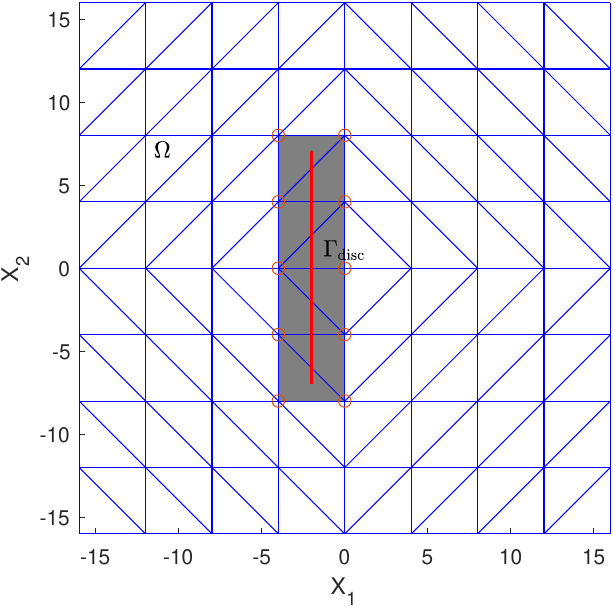}}
    \hfill
    \subfloat[]{\includegraphics[width=0.55\textwidth]{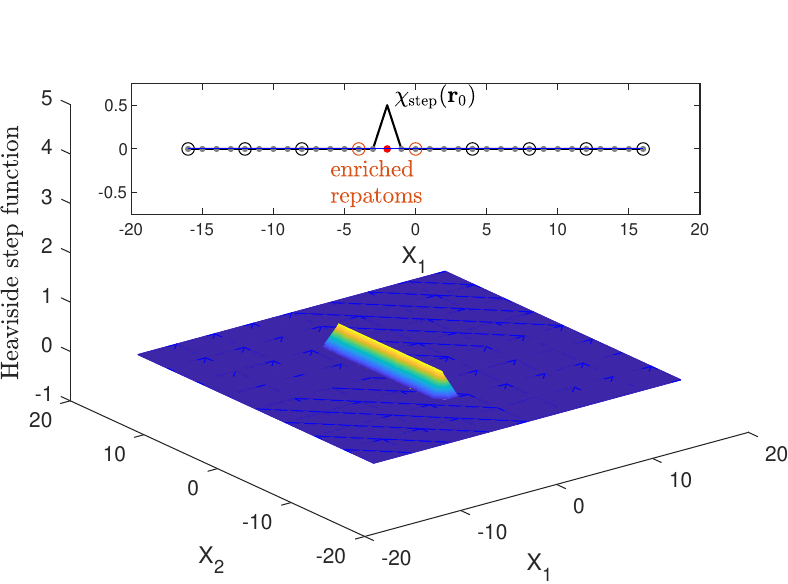}}
    \caption{(a) Example with a fiber as a discontinuity $\Gamma_{\mathrm{disc}}$ embedded in a domain $\Omega$, together with the enriched elements (dark grey) and repatoms (orange circles). (b) Step enrichment function for a fiber with a cross-section of $\chi_{\mathrm{step}}$ visualized as a 2D curve along the $X_1$-axis and for $X_2=0$ in the inset. The inset include the repatoms as open circles, the atoms as grey dots as well as the atom on the interface indicated as red dots.}
    \label{fig:SignDistFuncFiber}
\end{figure}

Beside weak discontinuities with an area domain such as the square inclusion with $\Omega_{2}$ in Figure \ref{fig:SignDistFunc}a, fibers are considered as weak discontinuities as well, and included by an enrichment function into the interpolation. The domain of the weak discontinuity collapses to the line $\Gamma_{\mathrm{disc}}$ (Figure \ref{fig:SignDistFuncFiber}a). Radtke et al. \cite{Radtke2010,Radtke2010b,Radtke2011} as well as Pike and Oskay \cite{Pike2015} have applied the Heaviside step function
\begin{equation}
    \label{eq:enrichment_step}
    \chi_{\mathrm{step}}(\mathbf{r}_0) = 
    \begin{cases}
        +0.5, & \psi(\mathbf{r}_0) = 0,\\
        0, & \psi(\mathbf{r}_0) > 0,
    \end{cases}
\end{equation}
for such weak discontinuities, which is derived from the signed distance function $\psi$ and characterized by having the value of $+0.5$ along $\Gamma_{\mathrm{disc}}$ and zero otherwise (Figure \ref{fig:SignDistFuncFiber}b).  


Independently of the enrichment strategy, the position vector of the $n_{\mathrm{ato}}$ atoms  
\begin{equation}
    \widehat{\mathbf{r}} = \boldsymbol{\Phi}\widehat{\mathbf{g}}
    \label{eq:interpoaltion}
\end{equation}
can be expressed by the interpolation matrix $\boldsymbol{\Phi}$ as well as the single column matrix of the generalized degrees of freedom 
\begin{equation}
    \widehat{\mathbf{g}} = [\widehat{\mathbf{r}}_{\mathrm{rep}},\widehat{\mathbf{g}}^{\star}_{1}, \ldots ,\widehat{\mathbf{g}}^{\star}_{n^{\star}}]^{\mathrm{T}}
\end{equation} 
containing the standard DOFs at the repatoms $\widehat{\mathbf{r}}_{\mathrm{rep}}$ as well as the enriched ones $\widehat{\mathbf{g}}^{\star}$ of the index set $N^{\star}$. The interpolation matrix $\boldsymbol{\Phi}$ in Eq. (\ref{eq:interpoaltion}) is formed in two steps where the first part consists of the standard FE interpolation 
\begin{equation}
    (\Phi_{\mathrm{FE}})_{(2\alpha-1)(2j-1)} = 
    \begin{cases}
        \phi_{\beta_j}(\mathbf{r}_0^{\alpha}), & \mathrm{for}\; \alpha \in N_{\mathrm{ato}}, \beta_j \in N_{\mathrm{rep}}, j=1,\ldots,n_{\mathrm{rep}} \\
        0, & \mathrm{otherwise}
    \end{cases}
\end{equation}
and the second part  
\begin{equation}
    (\Phi^{\star})_{(2\alpha-1)(2j-1)} =
    \begin{cases}
        \phi_j^{\star}(\mathbf{r}_0^{\alpha}), & \mathrm{for}\; \alpha \in N_{\mathrm{ato}}, j=1,\ldots,n^{\star} \\
        0, & \mathrm{otherwise}
    \end{cases}
\end{equation}
results from the enrichment in Eq. (\ref{eq:interpoaltion_enrichment}). Both are combined into the resulting interpolation matrix
\begin{equation}
    \boldsymbol{\Phi} = [\boldsymbol{\Phi}_{\mathrm{FE}}, \boldsymbol{\Phi}^{\star}]
\end{equation}
and applied in Eq. (\ref{eq:interpoaltion}).

\subsection{QC Reduction by Summation Rule}
The second step to reduce the computational effort in the QC method is related to summation rules. Hereafter, the potential energy is approximated by using a first-order interaction based summation rule, i.e.
\begin{equation}
    \Pi^k(\widehat{\mathbf{r}}) \approx \widehat{\Pi}^k(\widehat{\mathbf{r}}) = \sum 
    \limits_{\alpha \beta \in S^{\mathrm{sam}}_{\mathrm{int}}} \omega_{\alpha\beta}  \pi^k_{\alpha\beta} (\widehat{\mathbf{r}}).
\end{equation}
Four interactions, one for each orientation (\SI{0}{\degree}, \SI{90}{\degree}, $\pm$\SI{45}{\degree}), are selected inside an element. The potential energy of the selected interaction $\pi^k_{\alpha\beta}$, connecting atoms $\alpha$ and $\beta$, are multiplied with a weight factor $\omega_{\alpha\beta}$ to represent the potential energy of all interactions in that element. These interactions and their weight factors are stored in an index set $S_{\mathrm{int}}^{\mathrm{sam}}$. By summing up the products of the energy $\pi^k_{\alpha\beta}$ with the weight factors $\omega_{\alpha\beta}$ in the set $S_{\mathrm{int}}^{\mathrm{sam}}$, the potential energy $\widehat{\Pi}^k(\widehat{\mathbf{r}})$ of the whole system is approximated. This summation rule was first suggested by Beex et al. \cite{Beex2015}. Here, the summation rule will be expanded for the extended QC with non-conforming meshes with Heaviside enrichment and therefore to elements cut by a material interface. In the following the summation rule for the standard QC and the expansion for the extended QC is described in detail. It is referred to as first-order summation rule, which is appropriate for a first-order interpolation but not suitable for higher order polynomials.

\begin{figure}[h]
   \centering
   \subfloat[Triangulation]{\includegraphics[width=0.45\textwidth]{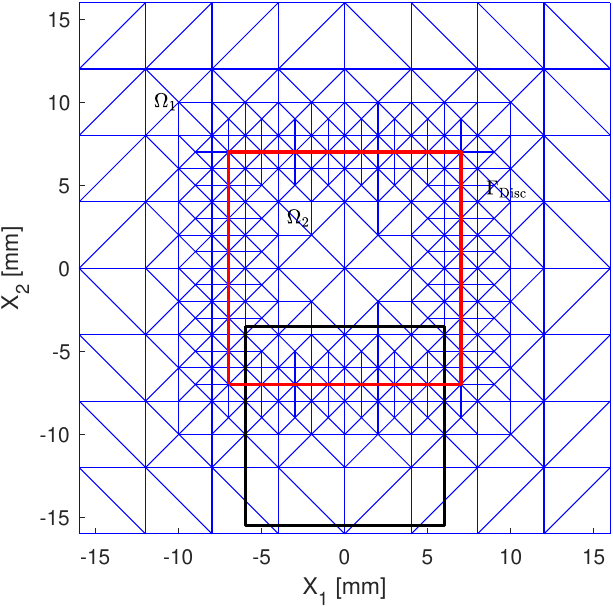}}
   \hfill
   \subfloat[Sampling Interactions]{\includegraphics[width=0.45\textwidth]{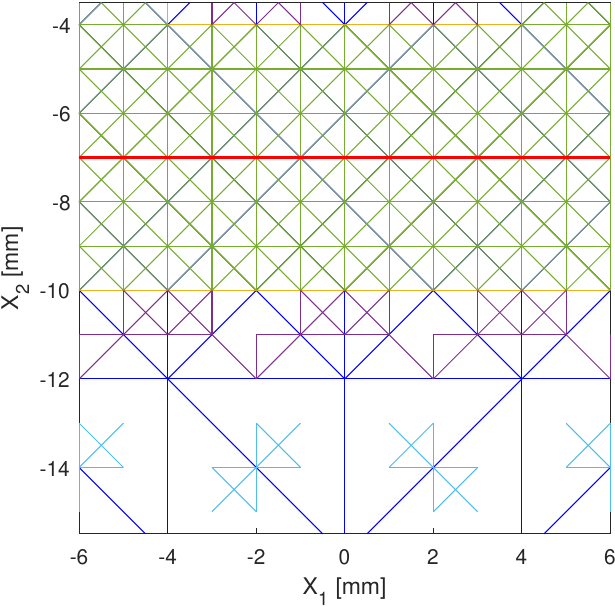}}
   \caption{(a) Triangulation of the standard QC along with the discontinuity (red) and the region of interest in a black frame together with (b) the magnification of the region of interest visualizing the sampling interactions for the triangles (bright blue: $\omega_{\alpha\beta}=8$, purple: $\omega_{\alpha\beta}=2$ or $\omega_{\alpha\beta}=4$, green: $\omega_{\alpha\beta}=1$ and yellow: $\omega_{\alpha\beta}=0.5$)}
   \label{fig:Triangulation_SamplingIntSQC}
\end{figure}

For the standard QC the selection procedure of the interactions as well as the calculation of the weight factors is outlined in Algorithm \ref{alg:SumRulStandQC}. 
\begin{algorithm}[tbh]
	\caption{First-order summation rule for standard QC} 
	\label{alg:SumRulStandQC}
	\begin{algorithmic}[1]
		\For {$i=1,\ldots,n_{\mathrm{T}}$}\Comment{Loop over all triangles}
		    \State Calculate area $A_i$ of current triangle based on lattice spacing $d$ 
		    \If {$A_i > \frac{1}{2}d^2$}
		        \State Find center point of the longest edge of the triangle $\mathbf{P}^i_{\mathrm{edge}}$
		        \State Find four interactions (one for each orientation) inside triangle at $\mathbf{P}^i_{\mathrm{edge}}$ 
		        \State Add four interactions to $S_{\mathrm{int}}^{\mathrm{sam}}$ set each with $\omega_{\alpha\beta} = A_i$
		    \Else
		        \For {$j=1,\ldots,n_{\mathrm{int}}$}\Comment{Loop over all interactions}
                    \State Compute center point $\mathbf{P}^j_{\mathrm{int}}$ of $j$-th interaction 
                    \State Barycentric coordinates of center point $\mathbf{P}^j_{\mathrm{int}}$
                    \If{$\mathbf{P}^j_{\mathrm{int}}$ inside the triangle}
                        \State Link is added to $S_{\mathrm{int}}^{\mathrm{sam}}$ set with $\omega_{\alpha\beta} = 1$
                    \ElsIf{$P_{\mathrm{int}}$ is on triangle edge}
                        \State Link is added to $S_{\mathrm{int}}^{\mathrm{sam}}$ set with $\omega_{\alpha\beta} = 0.5$
                    \EndIf
                \EndFor
		    \EndIf
		\EndFor
		\For {$i=2,\ldots,n_{\mathrm{sam}}$}
		    \State Sort sampling links in $S_{\mathrm{int}}^{\mathrm{sam}}$
		    \If{$S_{\mathrm{int}}^{\mathrm{sam}}(i-1) = S_{\mathrm{int}}^{\mathrm{sam}}(i)$}
		    \Comment{Check and remove for double occurrence of}
            \State \hspace{6.7cm} sampling links in ordered set $S_{\mathrm{int}}^{\mathrm{sam}}$
		        \State  $\omega_{\alpha\beta}^{i-1} = \omega_{\alpha\beta}^{i-1} + \omega_{\alpha\beta}^{i}$
		        \Comment{Sum the weight factors}
		    \EndIf
		\EndFor
	\end{algorithmic} 
\end{algorithm}
For meshes with right-angled triangles and an X-braced lattice used herein, the number of interactions and the weight factor scales with the area $A_i$ of the $i$-th element. Therefore, the weight factors for elements with edge size $h$ longer than the lattice spacing $d$ and $A_i > \frac{1}{2}d^2$ are assigned with the element area (Algorithm \ref{alg:SumRulStandQC}, lines
3–6). For the the fully resolved elements the weight factors are set to 1 or 0.5 to represent themselves, depending if they are located on the element edge or inside the element (Algorithm \ref{alg:SumRulStandQC}, lines
7–14). In Figure~\ref{fig:Triangulation_SamplingIntSQC} the selection procedure for the sampling interactions of Algorithm \ref{alg:SumRulStandQC} is illustrated. 
The model has a stiff inclusion embedded in a soft matrix, indicated by the red interface, and is fully resolved with triangles having the same size as the lattice spacing. With increasing distance from the discontinuity the element size grows towards the predefined edge length. In Figure~\ref{fig:Triangulation_SamplingIntSQC}b, the sampling interactions are shown enlarged. Close to the discontinuity all interactions are selected and added to the set $S_{\mathrm{int}}^{\mathrm{sam}}$ with a weight factor of $\omega_{\alpha\beta} = 1$. The elements with bright blue sampling interactions in Figure~\ref{fig:Triangulation_SamplingIntSQC}b have an area of $A_i = 8$ and each selected sampling interaction is assigned a weight factor of $\omega_{\alpha\beta} = A_i$. The same is done for the elements with purple sampling interactions between the fully resolved triangulation at the discontinuity with corresponding weight factors.

\begin{figure}[h]
   \centering
   \subfloat[Triangulation of standard QC]{\includegraphics[width=0.45\textwidth]{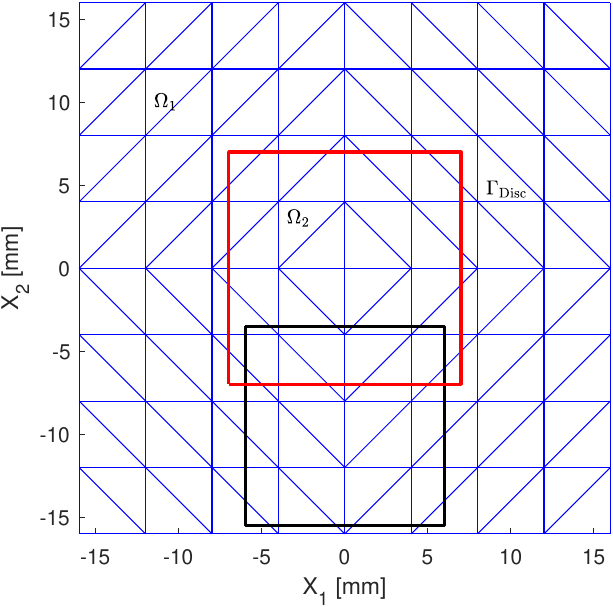}}
   \hfill
   \subfloat[Sampling Interactions]{\includegraphics[width=0.45\textwidth]{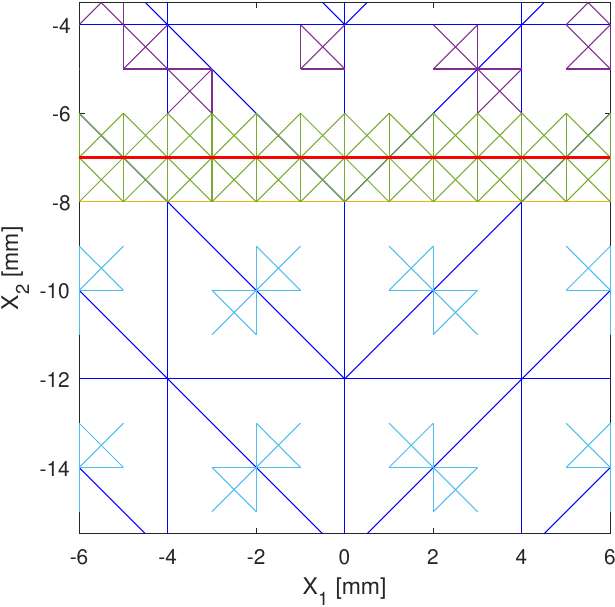}}
   \caption{(a) Triangulation of the extended QC along with the discontinuity (red) and the region of interest (highlighted as a black frame) together with (b) the magnification of the region of interest showing the sampling interactions for the triangles (bright blue: $\omega_{\alpha\beta}=4$, purple: $1<\omega_{\alpha\beta}<4$, green: $\omega_{\alpha\beta}=1$ and yellow: $\omega_{\alpha\beta}=0.5$)}
   \label{fig:Triangulation_SamplingIntXQC}
\end{figure}

For the extended QC the selection procedure of the sampling interactions of triangles which are not cut by the discontinuity is identical to the standard QC algorithm for the coarse mesh. The four sampling interactions are located at the center point of the longest edge of the element (Figure \ref{fig:Triangulation_SamplingIntXQC}) and the weight factors are determined through the area of the triangle (Algorithm \ref{alg:SumRulExtQC}, lines 26--30). For the triangles cut by the interface all interactions sharing one or two atoms with the $\Gamma_{\mathrm{disc}}$ are selected discretely. They are stored in the index set $S_{\mathrm{in}}^{\mathrm{disc}}$ which contains the interactions along the interface. The weight factor is thereby $\omega_{\alpha\beta}=1$ if the link is inside the triangle or $\omega_{\alpha\beta}=0.5$ if it is on a triangle edge.
\begin{algorithm}
	\caption{First-order summation rule for extended QC} 
	\label{alg:SumRulExtQC}
	\begin{algorithmic}[1]
	\For {$i=1,\ldots,n_{\mathrm{T}}$}\Comment{Loop over all triangles}
	    \If{distance function $\psi$ at least at one of the triangle atoms is 0} 
	        \For {$j=1,\ldots,n_{\mathrm{int}}$}\Comment{Loop over all interactions}
                \State Compute center point of interaction $\mathbf{P}^j_{\mathrm{int}}$ 
                \State Obtain Barycentric coordinates of center point $\mathbf{P}^j_{\mathrm{int}}$
                \If{$\mathbf{P}^j_{\mathrm{int}}$ inside the triangle}
                    \State Interaction is added to a set of links of $i$-th triangle $S_{\mathrm{int}}^{i}$ with $\omega_{\alpha\beta} = 1$
                \ElsIf{$\mathbf{P}^j_{\mathrm{int}}$ is on triangle edge}
                    \State Interaction is added to a set of links of $i$-th triangle $S_{\mathrm{int}}^{i}$ with $\omega_{\alpha\beta} = 0.5$
                \EndIf
                \If{$j$-th interaction shares at least one atom with the interface}
                    \State Set flag = 0
                \ElsIf{$j$-th interaction is inside inclusion}
                    \State Set flag = $-1$
                \ElsIf{$j$-th interaction is outside inclusion}
                    \State Set flag = $+1$
                \EndIf
            \EndFor 
            
            \For {$j=1,\ldots,n_{\mathrm{int Tri}}$}\Comment{Loop over all interactions of current triangle}
                \If{flag = 0}
                    \State Add to $S_{\mathrm{int}}^{\mathrm{disc}}$ discretely with $\omega_{\alpha\beta} = 1$ if interaction is inside the triangle \State and $\omega_{\alpha\beta} = 0.5$ if interaction is on triangle edge
                \ElsIf{flag = $-1$}
                    \State $\omega_{\alpha\beta}^{\mathrm{in}} = \omega_{\alpha\beta}^{\mathrm{in}} + \omega_{\alpha\beta}^j$
                \ElsIf{flag = $+1$}
                    \State $\omega_{\alpha\beta}^{\mathrm{out}} = \omega_{\alpha\beta}^{\mathrm{out}} + \omega_{\alpha\beta}^j$
                \EndIf
                \State Add last interaction with flag = $-1$ to $S_{\mathrm{int}}^{\mathrm{in}}$ and flag = $+1$ to $S_{\mathrm{int}}^{\mathrm{out}}$
                \State with $\{S_{\mathrm{int}}^{\mathrm{disc}}, S_{\mathrm{int}}^{\mathrm{in}}, S_{\mathrm{int}}^{\mathrm{out}}\} \subset S_{\mathrm{int}}^{\mathrm{sam}}$
            \EndFor
	    \Else \Comment {Triangle is not cut by the interface}
	        \State Find center point of the longest edge of the triangle $\mathbf{P}^i_{\mathrm{edge}}$
	        \State Find four interactions (one in each orientation) inside triangle 
	        \State and connected to $\mathbf{P}^i_{\mathrm{edge}}$
	        \State Add four interactions to set $S_{\mathrm{int}}^{\mathrm{sam}}$ each with $\omega_{\alpha\beta} = A_i$
	    \EndIf
	    
	\EndFor
	
	\For {$i=2,\ldots,n_{\mathrm{sam}}$}
        \State Sort sampling links in $S_{\mathrm{int}}^{\mathrm{sam}}$
		\If{$S_{\mathrm{int}}^{\mathrm{sam}}(i-1) = S_{\mathrm{int}}^{\mathrm{sam}}(i)$}
		    \Comment{Check and remove for double occurrence of}
            \State \hspace{6.7cm} sampling links in ordered set $S_{\mathrm{int}}^{\mathrm{sam}}$
		    \State  $\omega_{\alpha\beta}^{i-1} = \omega_{\alpha\beta}^{i-1} + \omega_{\alpha\beta}^{i}$
		    \Comment{Sum the weight factors}
		\EndIf
	\EndFor
	
	\For {$i=1,\ldots,n_{\mathrm{sam}}$}
		\If{$\omega_{\alpha\beta}^{i} = 0.5$ and $\theta_{\mathrm{int}}^{i} = \pm45^{\circ}$}
		    \State Find sampling interactions of neighbouring triangle the $i$-th link is shared with
		    \State Decrease the weight factor $\omega_{\alpha\beta}$ in that neighbouring triangle and direction by $0.5$ 
		    \State Set $\omega_{\alpha\beta}^{i} = 1$ for $i$-th interaction
		\EndIf
	\EndFor
    \end{algorithmic}
\end{algorithm}
Furthermore, for each of the two domains of the cut triangle one sampling interaction in all four orientations (\SI{0}{\degree}, \SI{90}{\degree} and $\pm$\SI{45}{\degree}) is selected. The links inside the inclusion are stored in the index set $S_{\mathrm{in}}^{\mathrm{int}}$ and the ones in the domain outside the inclusion are stored $S_{\mathrm{out}}^{\mathrm{int}}$ (Algorithm \ref{alg:SumRulExtQC}, lines 16--23). The three sets $S_{\mathrm{disc}}^{\mathrm{int}}$, $S_{\mathrm{in}}^{\mathrm{int}}$ and $S_{\mathrm{out}}^{\mathrm{int}}$ are all part of $S_{\mathrm{sam}}^{\mathrm{int}}$.
After leaving the for-loop the last four links in that domain are selected as sampling bonds (Algorithm \ref{alg:SumRulExtQC}, lines 24--25) and appear therefore at a corresponding locations in the triangle (purple in Figure \ref{fig:Triangulation_SamplingIntXQC}b). The second for--loop (lines 31--34 in Algorithm \ref{alg:SumRulExtQC}) checks for double occurrences of sampling links and is combining the weight factors. This ensures that a sampling link sharing at least one atom with the interface and which is in addition shared between two triangles (being on an edge) will be present only once in the set $S_{\mathrm{in}}^{\mathrm{int}}$ with a weight factor of $\omega_{\alpha\beta} = 1$ at the end of the procedure. It was observed that diagonal sampling interactions with weight factors $\omega_{\alpha\beta} = 0.5$ and connecting atoms in two different triangles lead to errors in the energy approximation. Therefore, the last step in Algorithm \ref{alg:SumRulExtQC} is to increase the weight factor to $\omega_{\alpha\beta} = 1$ and compensate for this error. At the same time the weight factor of the sampling interaction in the same direction in the neighbouring triangle (purple in Figure \ref{fig:Triangulation_SamplingIntXQC}b) is reduced by 0.5. These weight factors account for the exact number of interactions in every triangle as well as for the exact number of interactions in every domain of the cut triangles.

\section{Numerical Examples}
The QC method in its standard and extended version is applied to investigate an example with a single stiff circular inclusion (Section \ref{sec:SingleStiffInclusion}), a single fiber embedded in a soft matrix (Section \ref{sec:SingleFiber}) and a polished cross-section of a concrete sample with multiple aggregates and fibers. The focus is thereby on the convergence behavior of the displacement, elastic energy, homogenized stress and stiffness of the standard QC and its extended version. The full and the first-order interaction based summation rule are compared to each other for the three numerical examples. For all evaluations of the QC outcome the full solution of the lattice network is used as a reference.

\subsection{Single Stiff Inclusion}
\label{sec:SingleStiffInclusion}
\begin{figure}[h]
    \centering
    \includegraphics[width=0.35\textwidth]{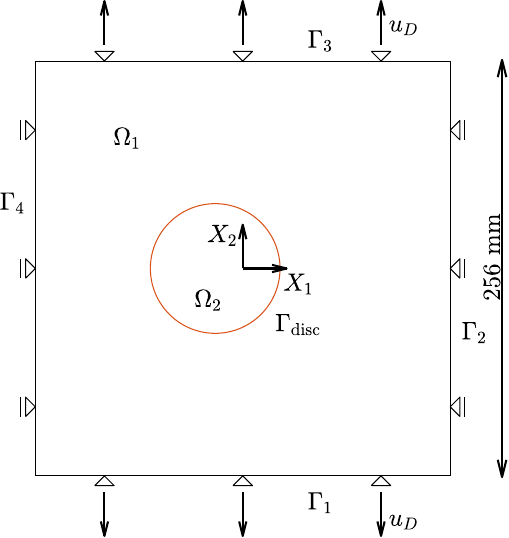}
    \caption{Numerical example with a single stiff circular inclusion with its center at $\mathbf{P}_\mathrm{cent} = (-17, 0)$ and a radius of $R = \SI{40}{mm}$ (red) and the applied boundary conditions including the prescribed displacement $u_D$ on $\Gamma_1$ and $\Gamma_3$.}
   \label{fig:Schematic_CircIncl_OC}
\end{figure}

The first numerical example is characterized by an inclusion with a radius of $R=\SI{40}{mm}$ and its center point at $\mathbf{P}_\mathrm{cent} = (-17,0)$, embedded in a square domain with a edge length of \SI{256}{mm}. The Young's modulus of the inclusions is ten times higher compared to the surrounding matrix. The underlying lattice has a spacing of $d=\SI{1}{mm}$ and is subject to tensile loading (Figure \ref{fig:Schematic_CircIncl_OC}). The loading is applied with the displacement $u_\mathrm{D}$ on the boundaries $\Gamma_1$ and $\Gamma_3$. All four boundaries are restricted in the $X_1$ direction while the nodes on $\Gamma_2$ and $\Gamma_4$ are free to move in the $X_2$ direction.  

\subsubsection{Convergence Behavior using the Full Summation Rule}
\label{sec:ConvergenceBehaviorFullSummationRule}
\begin{figure}[h]
   \centering
   \subfloat[standard QC]{\includegraphics[width=0.45\textwidth]{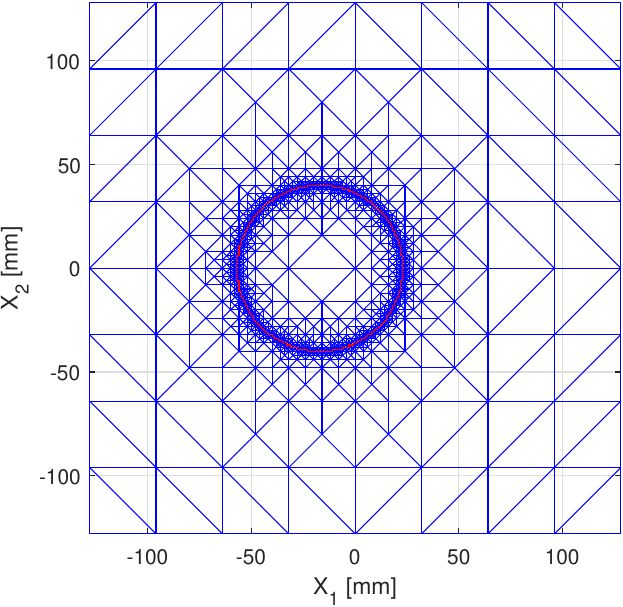}}
   \hfill
   \subfloat[extended QC]{\includegraphics[width=0.45\textwidth]{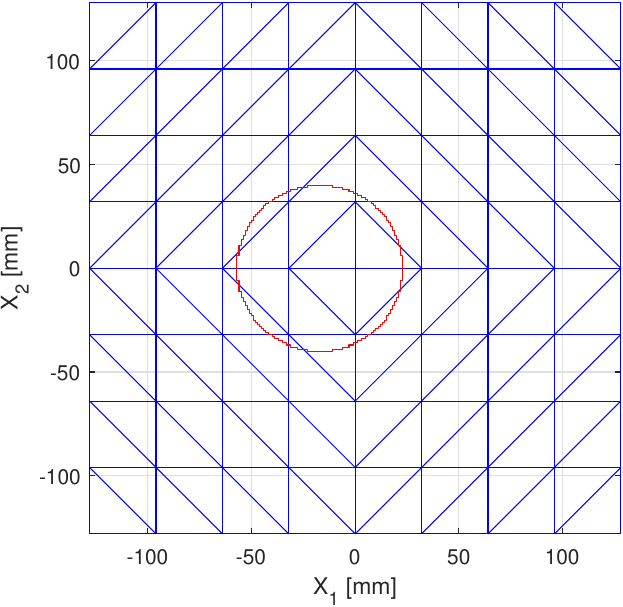}}
   \caption{Triangulation with an element size of \SI{32}{mm} of the (a) standard and (b) extended QC along with the circular inclusion in red}
   \label{fig:Triangulation_sQC_xQC}
\end{figure}

The example is simulated by using the full solution, the standard and extended QC Method. The extended QC includes four different enrichment strategies introduced in Section \ref{sec:Interpolation}. For both types of QC five different triangulations are used and the coarsest one is shown in Figure \ref{fig:Triangulation_sQC_xQC}.
The standard QC discretization is conforming and fully resolved along the discontinuity $\Gamma_{\mathrm{disc}}$ (red in Figure \ref{fig:Triangulation_sQC_xQC}) while the triangulation of the extended QC is characterized by a regular mesh. The discretization in Figure \ref{fig:Triangulation_sQC_xQC} is centrosymmetric and the uniform part of it has an edge length of \SI{32}{mm} which leads to approximately \SI{3200}{} DOFs for the triangulation of the standard and 200 for the extended QC, while the full solution has around \SI{133000}{} unknowns. Table \ref{tab:DOF} gives an overview of all numbers of DOFs with the corresponding triangle size. The full summation rule is applied for the standard and extended QC analyses and therefore the potential energy of the system is determined by visiting every interaction of the lattice. 
\begin{table}[h]
\centering
\caption{Element size of the triangulation together with the number of DOFs for the standard and extended QC using Heaviside enrichment.}
\begin{tabular}{lrrrrrr}
Element Size [mm] & 32 & 16 & 8 & 4 & 2 \\
\hline
DOF standard QC & \SI{3150}{} & \SI{3442}{} & \SI{4818}{} & \SI{10658}{} & \SI{34674}{} \\
DOF extended QC & \SI{194}{} & \SI{634}{} & \SI{2286}{} & \SI{8662}{} & \SI{33678}{}
\end{tabular}
\label{tab:DOF}
\end{table}

\begin{figure}[h]
   \centering
   \subfloat[Elastic Energy]{\includegraphics[width=0.45\textwidth]{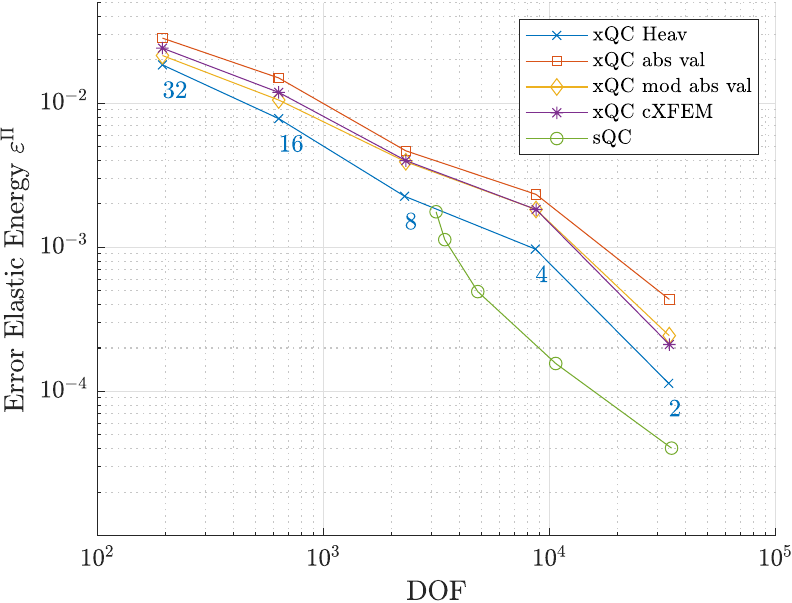}}
   \hfill
   \subfloat[Displacement]{\includegraphics[width=0.45\textwidth]{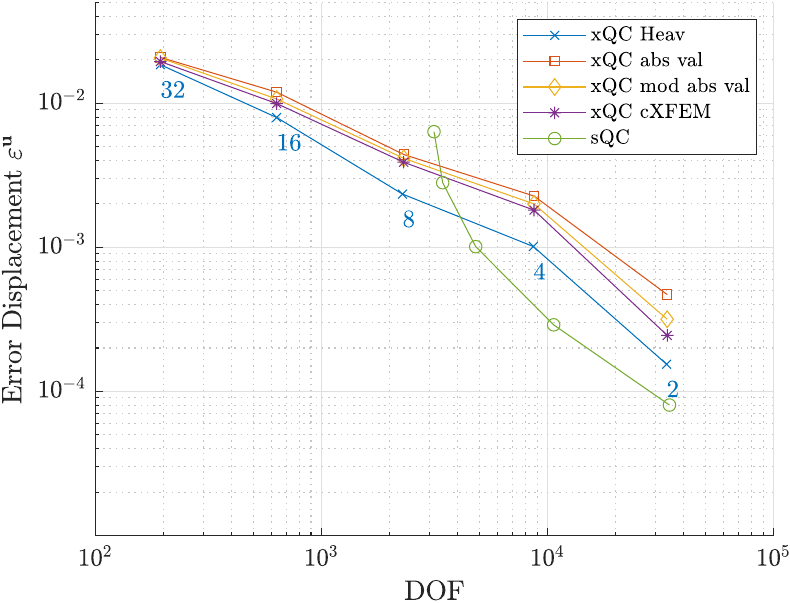}}
   \caption{(a) The relative error in elastic energy, (b) the relative error in displacement of the extended QC with the Heaviside (Heav), the absolute value (abs val), the modified absolute value (mod abs val) and the corrected XFEM (cXFEM) enrichment as well as the standard QC of the numerical example with a single stiff inclusion}
   \label{fig:ErEnDOF_ErDispDOF_CircIncl}
\end{figure}

The convergence behavior of the extended and standard QC is evaluated by the relative error in elastic energy and displacement magnitude. The relative error of the elastic energy
\begin{align}
    \varepsilon^{\Pi} = \dfrac{\lvert \Pi^{\mathrm{QC}}-\Pi^{\mathrm{FS}} \rvert}{\lvert \Pi^{\mathrm{FS}}\rvert}
    \label{eq:Error_Energy}
\end{align}
is determined from the difference between the QC analyses and the full solution, with regard to the absolute value of the elastic energy of the full solution $\lvert \Pi^{\mathrm{FS}}\rvert$. Moreover, the relative error of all displacements 
\begin{align}
    \varepsilon^{\mathbf{u}} = \dfrac{\lvert\lvert \textbf{u}^{\mathrm{QC}}-\textbf{u}^{\mathrm{FS}} \rvert\rvert_{L2}}{\lvert\lvert \textbf{u}^{\mathrm{FS}}\rvert\rvert_{L2} }
    \label{eq:Error_Disp}
\end{align}
is calculated, where $\mathbf{u} = \mathbf{r}-\mathbf{r}_0$. This leads to four curves for the extended QC with the Heaviside, absolute value, modified absolute value and corrected XFEM enrichment and one curve from the standard QC (Figure \ref{fig:ErEnDOF_ErDispDOF_CircIncl}). 

The error of the extended QC decreases exponentially with an increasing number of DOFs and all four curves are appearing as nearly straight lines in the double logarithmic diagrams. In addition, all four enrichment types result in a very similar outcome, whereby the Heaviside enrichment leads to the smallest and the absolute value enrichment to the largest error in elastic energy and displacement magnitude. The coarsest triangulation of the extended QC with 194 DOFs using the Heaviside enrichment has an error in elastic energy and displacement magnitude of approximately \SI{2}{\%} for both quantities. The same enrichment leads to the smallest error between \SI{0.01}{\%} and \SI{0.02}{\%} for both quantities using the finest discretization with 33678 DOFs. The error for the standard QC ranges between \SI{0.2}{\%} and \SI{4e-3}{\%} for the elastic energy and between \SI{0.65}{\%} and \SI{8e-3}{\%} for the displacement magnitude (Figures \ref{fig:ErEnDOF_ErDispDOF_CircIncl}a and b).

\begin{figure}[h]
   \centering
   \subfloat[Element Size 32 mm]{\includegraphics[width=0.33\textwidth]{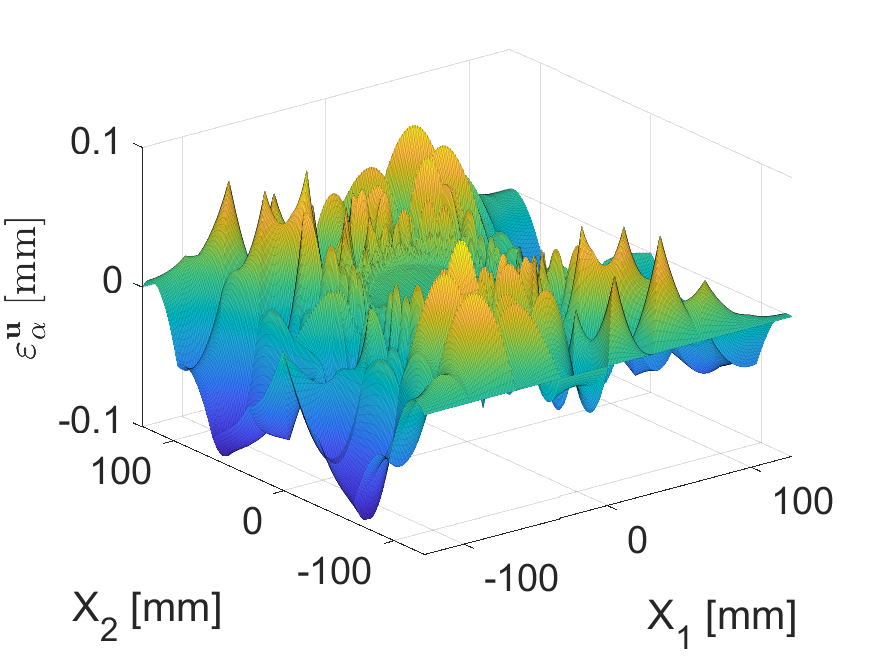}}
   \hfill
   \subfloat[Element Size 8 mm]{\includegraphics[width=0.33\textwidth]{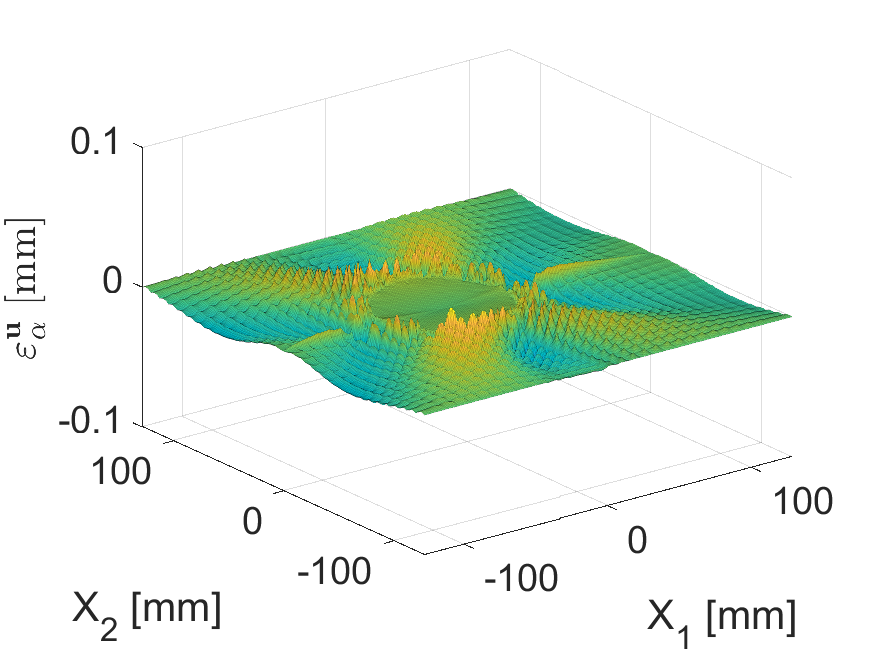}}
   \subfloat[Element Size 2 mm]{\includegraphics[width=0.33\textwidth]{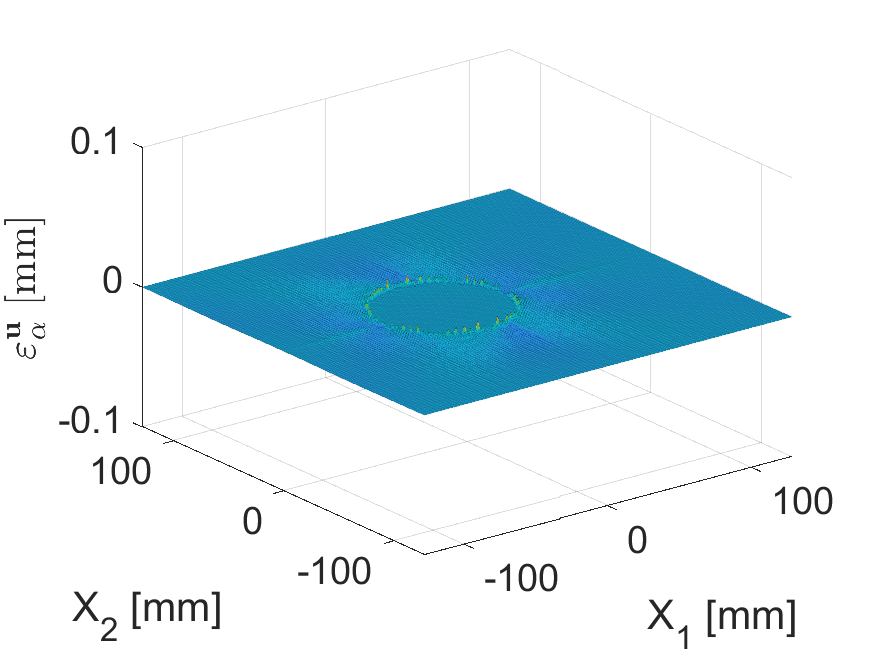}}
   \caption{The field plot of the local error in displacement magnitude of the standard QC for the numerical example with a single stiff inclusion}
   \label{fig:ErDispMag_CircIncl_sQC}
\end{figure}

\begin{figure}[h]
   \centering
   \subfloat[Element Size 32 mm]{\includegraphics[width=0.33\textwidth]{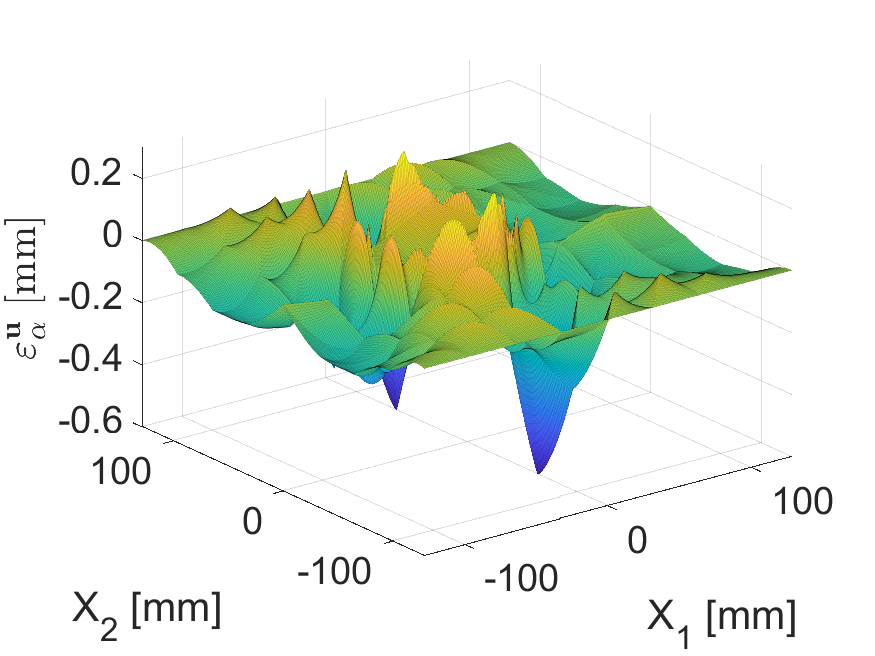}}
   \hfill
   \subfloat[Element Size 8 mm]{\includegraphics[width=0.33\textwidth]{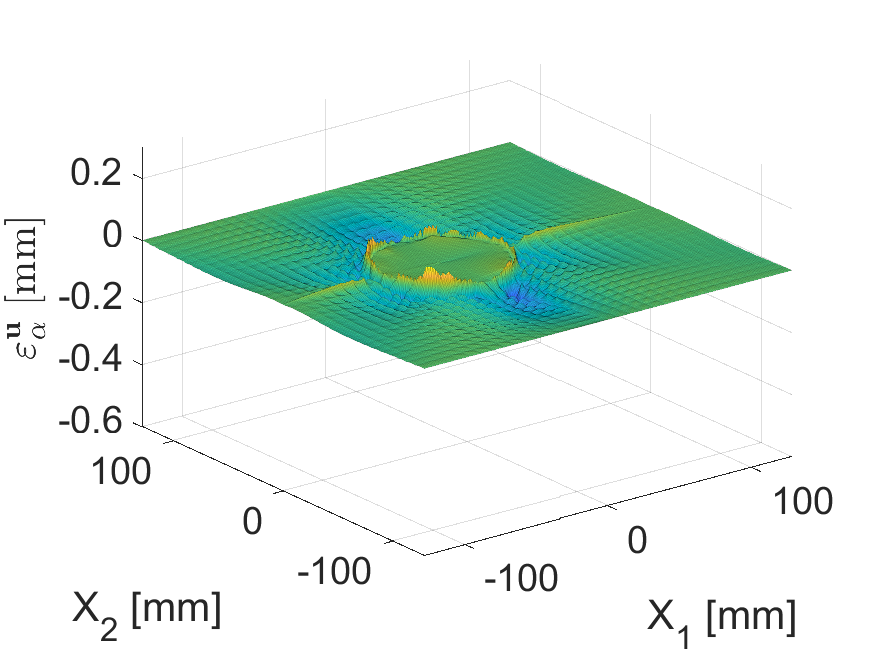}}
   \hfill
   \subfloat[Element Size 2 mm]{\includegraphics[width=0.33\textwidth]{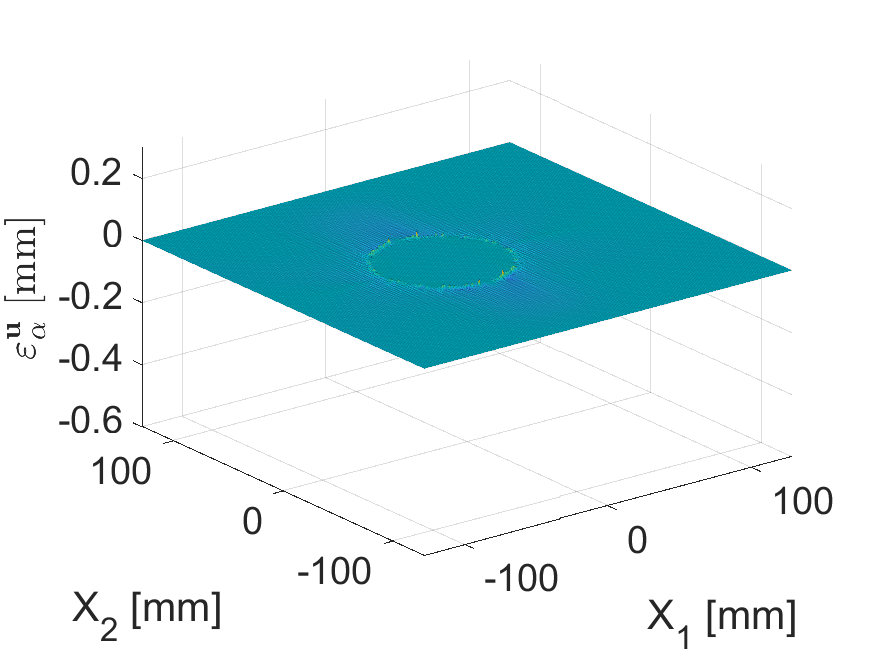}}
   \caption{The field plot of the local error in displacement magnitude of the extended QC with Heaviside enrichment for the numerical example with a single stiff inclusion.}
   \label{fig:ErDispMag_CircIncl_Heav}
\end{figure}

\begin{figure}[h]
   \centering
   \subfloat[Element Size 32 mm]{\includegraphics[width=0.33\textwidth]{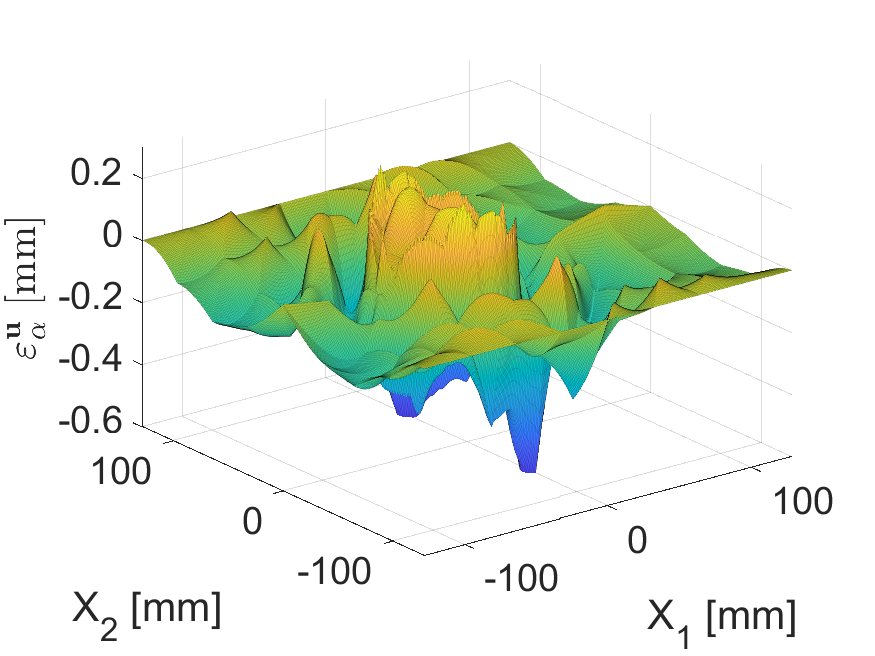}}
   \hfill
   \subfloat[Element Size 8 mm]{\includegraphics[width=0.33\textwidth]{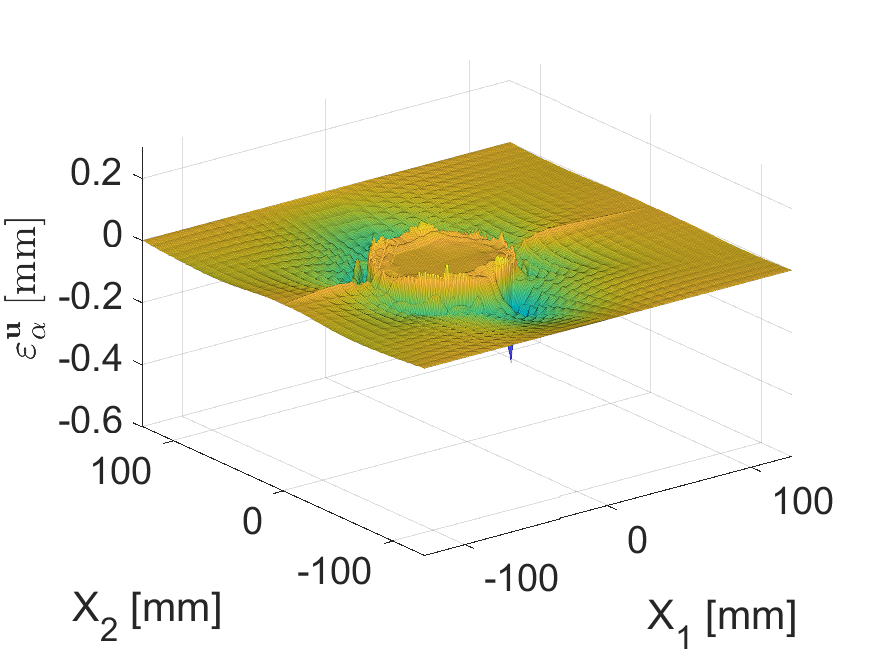}}
   \hfill
   \subfloat[Element Size 2 mm]{\includegraphics[width=0.33\textwidth]{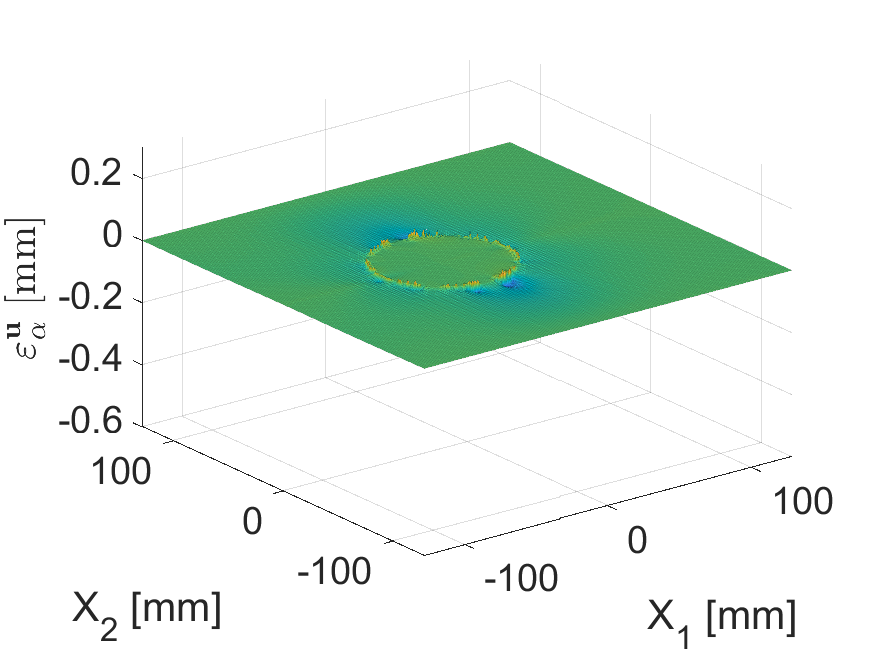}}
   \caption{The field plot of the local error in displacement magnitude of the extended QC with absolute value enrichment for the numerical example with a single stiff inclusion.}
   \label{fig:ErDispMag_CircIncl_abs}
\end{figure}

In Figures \ref{fig:ErDispMag_CircIncl_sQC}, \ref{fig:ErDispMag_CircIncl_Heav} and \ref{fig:ErDispMag_CircIncl_abs} field plots of the absolute error of the displacement magnitude are visualising the distribution of difference of both types compared to the full solution. Three out of the five discretizations are displayed, along with the extended QC incorporating Heaviside (Figure \ref{fig:ErDispMag_CircIncl_Heav}) and absolute value enrichment (Figure \ref{fig:ErDispMag_CircIncl_abs}). The absolute error of the displacement magnitude 
\begin{align}
    \boldsymbol{\varepsilon}^{\mathbf{u}}_{\alpha} = \lvert\lvert \textbf{u}^{\mathrm{QC}}_{\alpha}\rvert\rvert - \lvert\lvert\textbf{u}^{\mathrm{FS}}_{\alpha} \rvert\rvert, \qquad \quad \alpha = 1,\dots,n_{\mathrm{ato}},
\end{align}
is calculated for every atom $\alpha$ as the difference between the QC prediction and the reference result of the full solution. For the standard QC the error in displacement occurs throughout the whole domain of the compliant matrix, but is visibly smaller inside the stiff inclusion (Figure \ref{fig:ErDispMag_CircIncl_sQC}). In case of the extended QC the inclusion domain protrudes in the error surface and especially the interface shows local maxima. The surrounding matrix domain is characterized by mostly negative error values and indicates a too stiff prediction of the matrix behaviour caused by coarse interpolation (Figure \ref{fig:ErDispMag_CircIncl_Heav} and \ref{fig:ErDispMag_CircIncl_abs}).  

\subsubsection{Convergence of the First-Order Summation Rule}
\begin{figure}[htb]
   \centering
   \subfloat[Elastic Energy]{\includegraphics[width=0.45\textwidth]{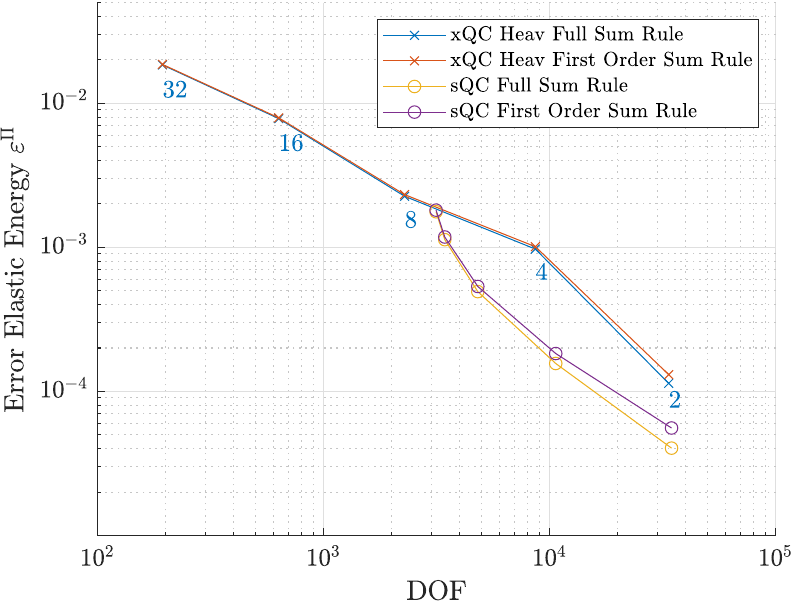}}
   \hfill
   \subfloat[Displacement]{\includegraphics[width=0.45\textwidth]{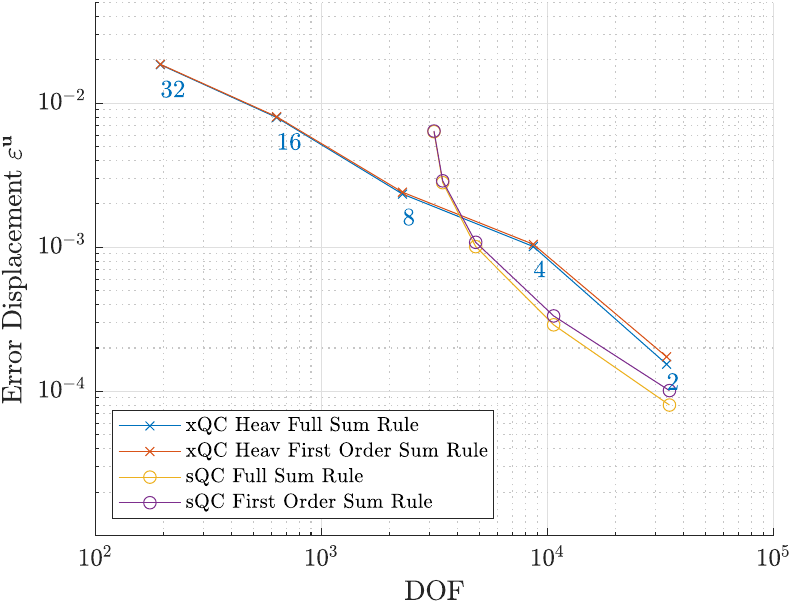}}
   \caption{(a) The relative error in elastic energy, (b) the relative error in displacement of the extended QC using the Heaviside enrichment (Heav) and the standard QC with full and first-order summation rule of the numerical example with a single stiff circular inclusion.}
   \label{fig:Error_EnRF_FOSumRule}
\end{figure}

The first-order summation rule is characterized by selecting only one sampling interaction in all four orientations inside a triangle or domain on both sides of the discontinuity. The selection procedure for the standard and extended QC is explained in Section \ref{alg:SumRulExtQC} and visualized in Figures \ref{fig:Triangulation_SamplingIntSQC} and \ref{fig:Triangulation_SamplingIntXQC}. To quantify the influence of the first-order summation rule relative to the full summation rule on the accuracy of the analyses results both approaches are compared. Since the absolute value, the modified absolute value as well as the corrected XFEM enrichment types need higher order summation rules the comparison is limited to the enriched interpolation with the Heaviside function. For all studied mesh sizes the Heaviside enrichment led to smallest error in elastic energy and displacement using the full summation rule (Figure \ref{fig:ErEnDOF_ErDispDOF_CircIncl}) and justifies this limitation. The full and first-order summation rule leads to very similar results of the relative error of the elastic energy and displacement magnitude for the standard and extended QC (Figure \ref{fig:Error_EnRF_FOSumRule}). The difference of relative error for the elastic energy and displacement magnitude between the summation rules is for all triangulations smaller than \SI{0.015}{\%} and decreases with element size. On the other hand, the number of sampling interactions and therefore the computationally effort is substantially smaller. The full summation rule uses all \SI{262656}{} interactions while the first-order summation rule reduces that number to \SI{3.7}{\%} to \SI{50.9}{\%} for the standard QC and \SI{1}{\%} to \SI{50.5}{\%} for the extended QC, depending on the element size (Table \ref{tab:Samp_Interactions_CircIncl}).

\begin{table}[h]
\centering
\caption{Number of sampling interactions for each element size of the triangulation for the standard and extended QC using the Heaviside enrichment.}
\begin{tabular}{lrrrrrr}
Element Size [mm] & 32 & 16 & 8 & 4 & 2 \\
\hline
Sampling interactions standard QC & \SI{9680}{} & \SI{10720}{} & \SI{15968}{} & \SI{38816}{} & \SI{133856}{} \\
Sampling interactions extended QC & \SI{2628}{} & \SI{4212}{} & \SI{10362}{} & \SI{34944}{} & \SI{132690}{}
\end{tabular}
\label{tab:Samp_Interactions_CircIncl}
\end{table}

\subsection{Single Fiber}
\label{sec:SingleFiber}
\begin{figure}[tbh]
   \centering
   \includegraphics[width=0.35\textwidth]{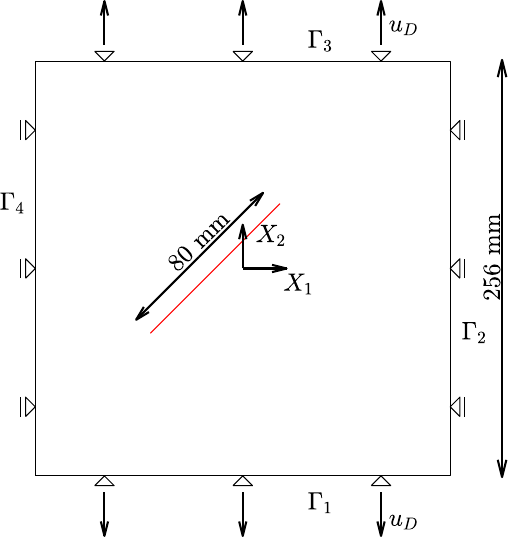}
   \caption{Numerical example with a single fiber with its center at $\mathbf{P}_\mathrm{cent} = (-17, 0)$ and a length of \SI{80}{mm} (red) and the applied boundary conditions including the prescribed displacement $u_D$ on $\Gamma_1$ and $\Gamma_3$}
   \label{fig:Schematic_Fiber}
\end{figure}

The standard and extended QC is next applied to an example of a single fiber as weak discontinuity embedded in a compliant matrix. The fibers Young's modulus is 100 times higher than that of the matrix, and the fiber has its center at $\mathbf{P}_{\mathrm{cent}} = (-17, 0)$ oriented under \SI{45}{\degree} (Figure \ref{fig:Schematic_Fiber}). 
It has a length of \SI{80}{mm} and the entire model is \SI{256}{mm} in width and height. The boundary conditions are the same as in the previous numerical example with a prescribed displacement $u_\mathrm{D}$ on $\Gamma_1$ and $\Gamma_3$ in the $X_2$ direction and a restricted displacement in the $X_1$ direction on all four boundaries. The enrichment for the extended QC is based on the step function in Eq. (\ref{eq:enrichment_step}).

\begin{figure}[h]
   \centering
   \subfloat[Elastic Energy]{\includegraphics[width=0.5\textwidth]{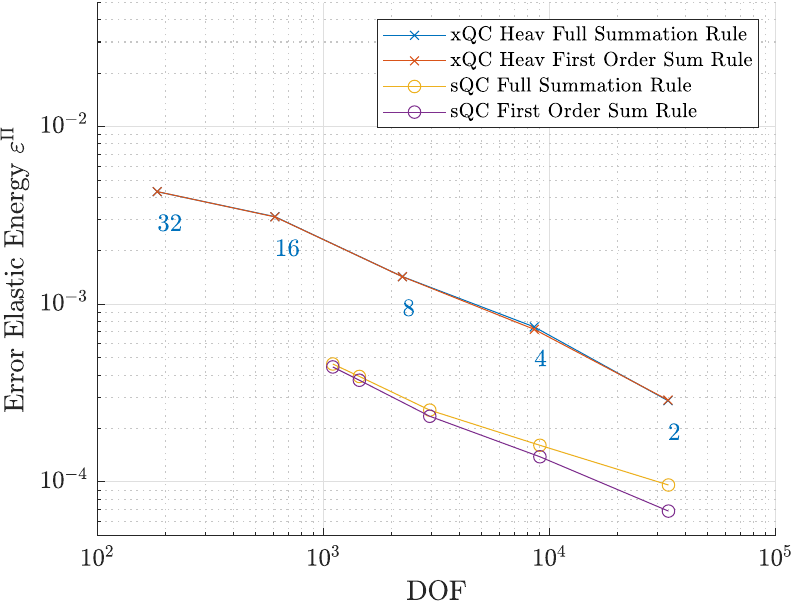}}
   \hfill
   \subfloat[Displacement Magnitude]{\includegraphics[width=0.5\textwidth]{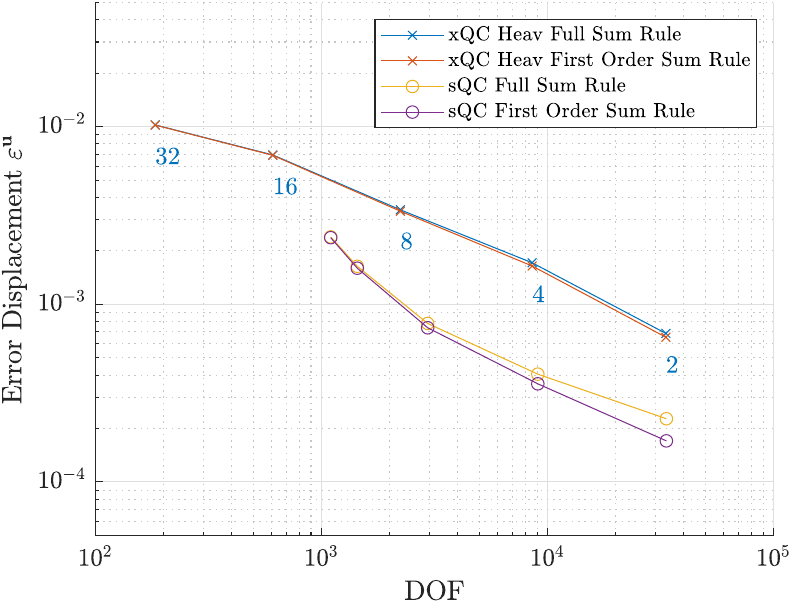}}
   \caption{(a) The relative error in elastic energy and the (b) relative error in displacement using the extended QC with the Heaviside enrichment (Heav) and the standard QC with full and first-order summation rule of the fiber example.}
   \label{fig:Error_EnRF_Fiber}
\end{figure}

The convergence behavior of the standard and extended QC for the relative error of the elastic energy and the displacement magnitude according to Eq. (\ref{eq:Error_Energy}) and (\ref{eq:Error_Disp}) is visualised in Figure \ref{fig:Error_EnRF_Fiber} using the full solution as a reference. The standard QC leads, for all triangulations and for both quantities, to a higher accuracy compared to the extended QC. 
Nevertheless, the extended QC is predicting the elastic energy with a relative error smaller than \SI{4.3E-1}{\%} while the relative error for the displacement magnitude is below \SI{1}{\%} (Figure \ref{fig:Error_EnRF_Fiber}), which is well acceptable. The first-order summation rule has a marginal influence on the outcome for the two QC approaches and the largest difference is the relative error of displacement magnitude \SI{7E-3}{\%} of the standard QC.

\subsection{Multiple Inclusions and Fibers}
\begin{figure}[h]
   \centering
   \subfloat[Concrete cross-section]{\includegraphics[width=0.45\textwidth]{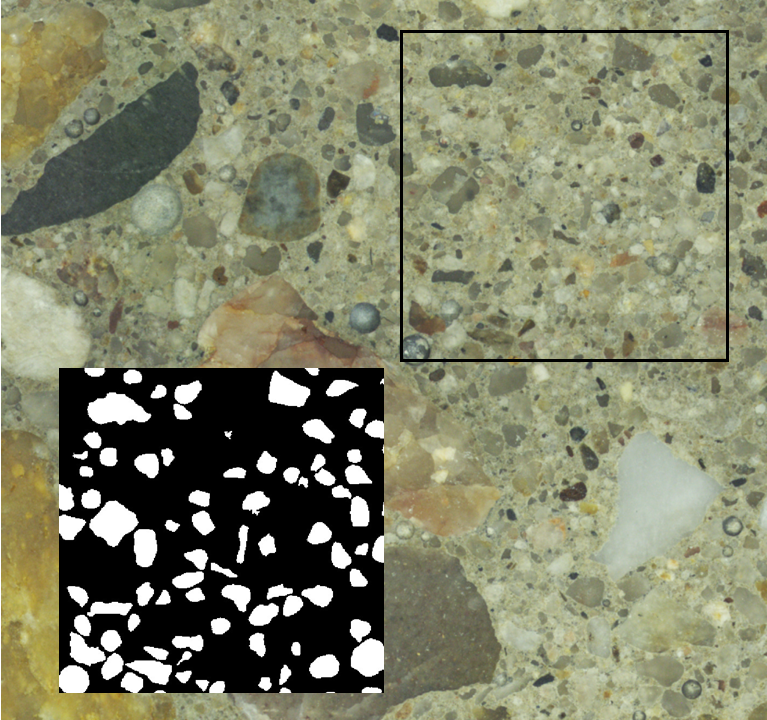}}
   \hfill
   \subfloat[Schematic with PBC]{\includegraphics[width=0.45\textwidth]{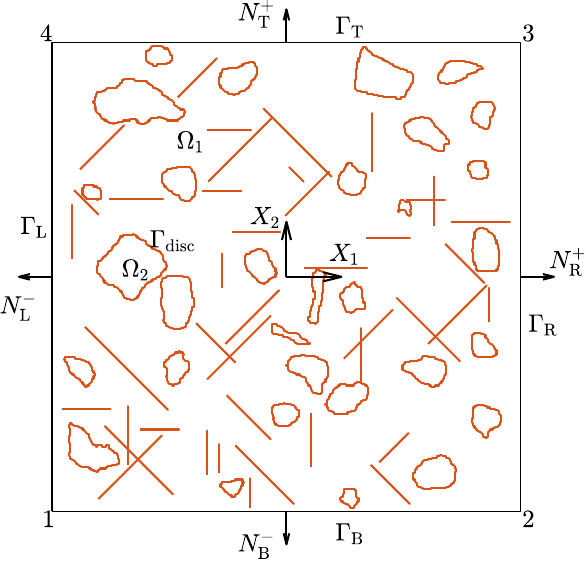}}
   \caption{(a) Concrete cross-section together with the segmentation of the region of interested indicated by the frame. (b) Schematic of the example with 31 inclusions and 42 fibers with periodic boundary conditions}
   \label{fig:Schematic_PBC}
\end{figure}

The third example is representing a realistic and complex microstructure, where the inclusion geometry is taken from a segmented concrete cross-section representing aggregates embedded in a cement matrix (Figure \ref{fig:Schematic_PBC}a). From the segmented cross-section 31 inclusions are selected for the QC model and 42 fibers are added (Figure \ref{fig:Schematic_PBC}b). The standard and extended QC is used to study the convergence behavior of the elastic energy, the displacement magnitude as well as homogenized material parameters. The Young's modulus of the matrix is $\SI{1}{MPa}$, inclusions have $\SI{10}{MPa}$ and the fibers have $\SI{100}{MPa}$. The square domain size is $\SI{384}{mm}$ and the lattice spacing is $d=\SI{1}{mm}$.

\begin{figure}[h]
   \centering
   \subfloat[Standard QC]{\includegraphics[width=0.45\textwidth]{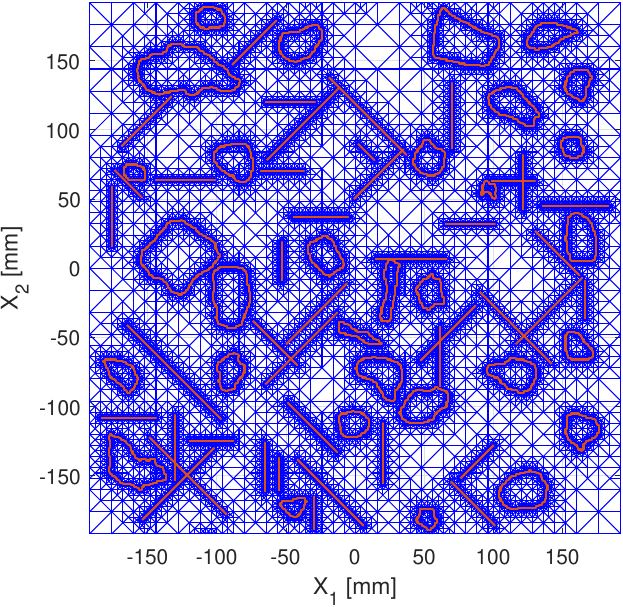}}
   \hfill
   \subfloat[Extended QC]{\includegraphics[width=0.45\textwidth]{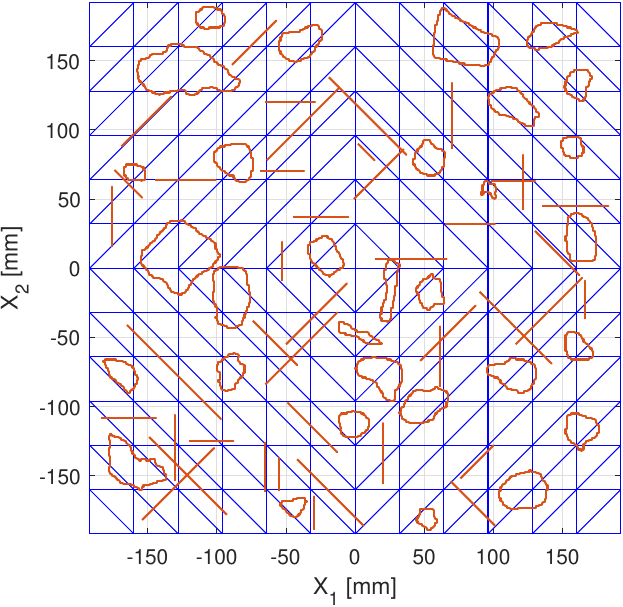}}
   \caption{Triangulation of (a) the standard and (b) extended QC with multiple inclusions and fibers.}
   \label{fig:31InclConc42Fib}
\end{figure}

For the standard QC the automatic refinement algorithm starts out with the same element size as in the previous two examples, but due to the high number of interfaces and the requirement to resolve them fully leads to relatively fine mesh throughout the whole domain (Figure \ref{fig:31InclConc42Fib}). The four coarsest triangulation have therefore a similar number of DOFs and only the finest triangulation with an element size of \SI{2}{mm} has a substantially different amount of unknowns (Table \ref{tab:DOF_31InclConc42Fib}). For the extended QC the inclusions and fibers have two separate level set functions and enrichments as described in Section \ref{sec:Interpolation}. For the inclusions the Heaviside sign function according to Eq. (\ref{eq:enrichment_Heav}) and for the fibers the Heaviside step function in Eq. (\ref{eq:enrichment_step}) is used. The triangulations for the extended QC range between \SI{800}{} and \SI{82000}{} DOFs, the standard QC has between \SI{50000}{} and \SI{99000}{} DOFs (Table \ref{tab:DOF_31InclConc42Fib}), while the full solution leads to approximately \SI{296000}{} unknowns.

Previous studies \cite{Tran20111436,Hiriyur2011257} have shown that in densely packed inclusions the XFEM with one level set function representing all inclusions can lead to artefacts. The enriched DOF is thereby unable to account for two different interfaces in neighbouring elements. A multi level set approach, which lead to enriched DOF to account for only one interface, can avoid this artefacts and further improve the convergence of effective elastic properties. In this study we limit ourselves to two level set function, one representing inclusions and one for fibers, although such extensions would be straight forward.

\begin{table}[h]
\centering
\caption{Element size of the triangulation together with the number of DOFs for the standard and extended QC}
\begin{tabular}{lrrrrrr}
Element Size in mm & 32 & 16 & 8 & 4 & 2 \\
\hline
DOF standard QC & \SI{50962}{} & \SI{50964}{} & \SI{51470}{} & \SI{57920}{} & \SI{98868}{} \\
DOF extended QC & \SI{842}{} & \SI{2398}{} & \SI{7120}{} & \SI{23134}{} & \SI{81818}{}
\end{tabular}
\label{tab:DOF_31InclConc42Fib}
\end{table}

\subsubsection{Periodic Boundary Conditions}
For such complex microstructures homogenized material parameters are of interest and therefore the homogenized first Piola-Kirchhoff stress tensor $\mathbf{P}$ as well as the the homogenized stiffness tensor $\mathbf{D}$ are determined, and periodic boundary conditions are applied to the Representative Volume Element (RVE) in Figure~\ref{fig:Schematic_PBC}b. 
The periodicity conditions
\begin{equation}
    \mathbf{r}_{\mathrm{rep}}^{+} - \mathbf{r}_{\mathrm{rep}}^{-} = \mathbf{F} (\mathbf{r}_{0 , \, \mathrm{rep}}^{+} - \mathbf{r}_{0  , \, \mathrm{rep}}^{-})
    \label{eq:period_cond}
\end{equation}
are used with $\mathbf{r}_{\mathrm{rep}}^{+}$ being the repatom locations on the boundaries $\Gamma_{\mathrm{R}}$ and $\Gamma_{\mathrm{T}}$ with normal vectors $N^{+}$ and $\mathbf{r}_{\mathrm{rep}}^{-}$ the repatom locations on $\Gamma_{\mathrm{B}}$ and $\Gamma_{\mathrm{L}}$ with $N^{-}$ (Figure \ref{fig:Schematic_PBC}b). The indices indicate bottom, top, left and right part of the boundary. The current location of the repatoms $\mathbf{r}_{\mathrm{rep}}^{+/-}$ along the boundaries is calculated by the macroscopic deformation gradient $\mathbf{F}$ and the location of the repatoms in the initial configuration $\mathbf{r}_{0 , \,  \mathrm{rep}}^{+/-}$. Here, a deformation gradient 
\begin{equation}
    \mathbf{F} = 
    \begin{bmatrix}
    1.06 & 0 \\
    0 &  1
    \end{bmatrix}
\end{equation}
leading to a tensile loading in the $X_1$ direction is applied. For completeness, the details on implementing the boundary conditions and extracting homogenized quantities are presented in \hyperref[AppendixA]{Appendix A}.

\subsubsection{Convergence Behavior}
\begin{figure}[h]
   \centering
   \subfloat[Elastic Energy]{\includegraphics[width=0.45\textwidth]{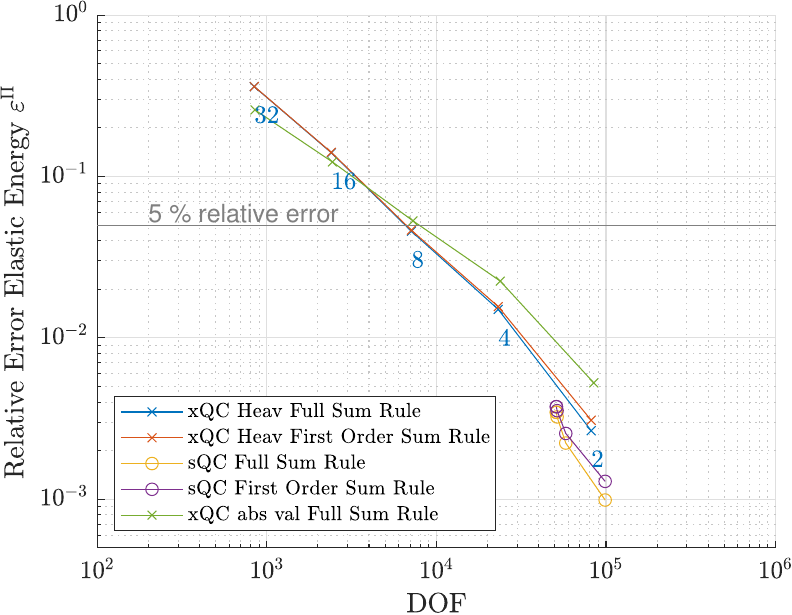}}
   \hfill
   \subfloat[Displacement]{\includegraphics[width=0.45\textwidth]{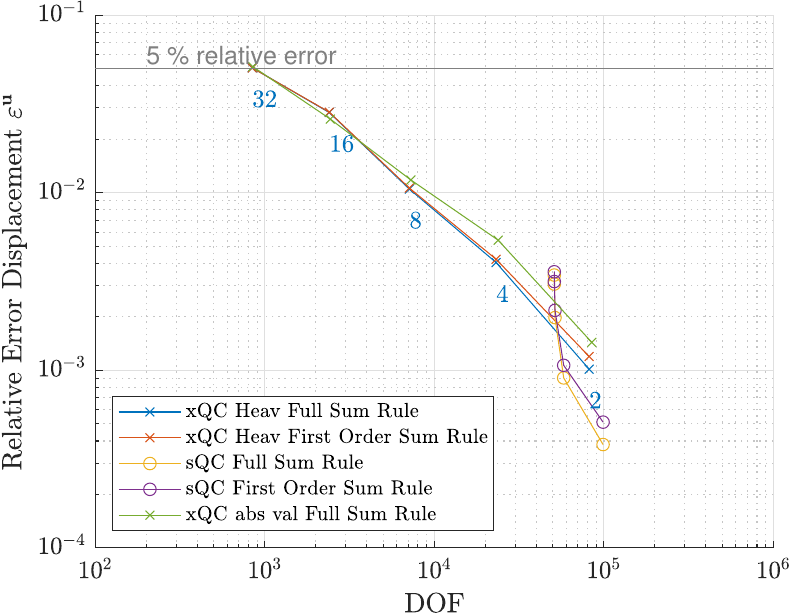}}
   \caption{The relative error of (a) the elastic energy and (b) the displacement magnitude of the multi inclusion and fiber example using the extended QC with the Heaviside (Heav) and the absolute value (abs val) enrichment and the standard QC with full and first-order summation rule}
   \label{fig:31InclConc42Fib_ErDispEn}
\end{figure}

For the standard and extended QC, using the full and first-order summation rule, the convergence behavior of the elastic energy and the displacement magnitude are determined according to Eqs. (\ref{eq:Error_Energy}) and (\ref{eq:Error_Disp}) (Figure \ref{fig:31InclConc42Fib_ErDispEn}). 
The relative error of the elastic energy using the extended QC with Heaviside enrichment is between \SI{36}{\%} and \SI{0.3}{\%} and is therefore more than one order of magnitude higher compared to the single inclusion example (Figure \ref{fig:Error_EnRF_FOSumRule}b). The prediction of the standard QC leads to relative errors between \SI{0.4}{\%} and \SI{0.1}{\%} of the elastic energy (Figure \ref{fig:31InclConc42Fib_ErDispEn}a). The relative error of the displacement magnitude ranges between \SI{5}{\%} and \SI{0.1}{\%}, while the standard QC leads to errors smaller than \SI{0.35}{\%}. In addition, analyses using the absolute value enrichment with full summation rule are carried out. They are slightly more accurate for the two coarser triangulations of the elastic energy, while the Heaviside enrichment is more accurate for finer discretizations for the elastic energy and the displacement (Figure \ref{fig:31InclConc42Fib_ErDispEn}). The differences between the two summation rules are similar to the two previous examples and are negligible. The largest deviation between the summation rules for both quantities is \SI{0.08}{\%} and occurs in the elastic energy for the coarsest triangulation of the extended QC (Figure \ref{fig:31InclConc42Fib_ErDispEn}a). The number of sampling interactions is reduced by the first-order summation rule from \SI{590592}{} to a range between \SI{26.3}{\%} and \SI{58.3}{\%} for the standard QC and to \SI{6.1}{\%} and \SI{55}{\%} for the extended QC depending on the element size of the mesh (Table~\ref{tab:Samp_Interactions_31InclConc42Fib}).
\begin{table}[h]
\centering
\caption{Number of sampling interactions for each element size of the triangulation for the standard and extended QC using Heaviside enrichment.}
\begin{tabular}{lrrrrrr}
Element Size [mm] & 32 & 16 & 8 & 4 & 2 \\
\hline
Sampling interactions standard QC & \SI{155234}{} & \SI{155242}{} & \SI{157074}{} & \SI{182202}{} & \SI{344490}{} \\
Sampling interactions extended QC & \SI{35901}{} & \SI{39967}{} & \SI{54585}{} & \SI{110239}{} & \SI{324929}{}
\end{tabular}
\label{tab:Samp_Interactions_31InclConc42Fib}
\end{table}

In addition, the homogenized stress $\mathbf{P}$ and stiffness tensor $\mathbf{D}$ are determined. The relative error of both quantities
\begin{equation}
    \varepsilon^{\mathbf{P}} = \dfrac{\lvert\lvert \mathbf{P}^{\mathrm{QC}}-\mathbf{P}^{\mathrm{FS}} \rvert\rvert_{\mathrm{F}}}{\lvert\lvert \mathbf{P}^{\mathrm{FS}} \rvert\rvert_{\mathrm{F}}}
    \qquad
    \mathrm{and}
    \qquad
    \varepsilon^{\mathbf{D}} = \dfrac{\lvert\lvert \mathbf{D}^{\mathrm{QC}}-\mathbf{D}^{\mathrm{FS}} \rvert\rvert_{\mathrm{F}}}{\lvert\lvert \mathbf{D}^{\mathrm{FS}} \rvert\rvert_{\mathrm{F}}}
\end{equation}
is determined by the Frobenius norm of the tensor of the QC analyses and the full solution. 
\begin{figure}[h]
   \centering
   \subfloat[Homogenized Stress]{\includegraphics[width=0.45\textwidth]{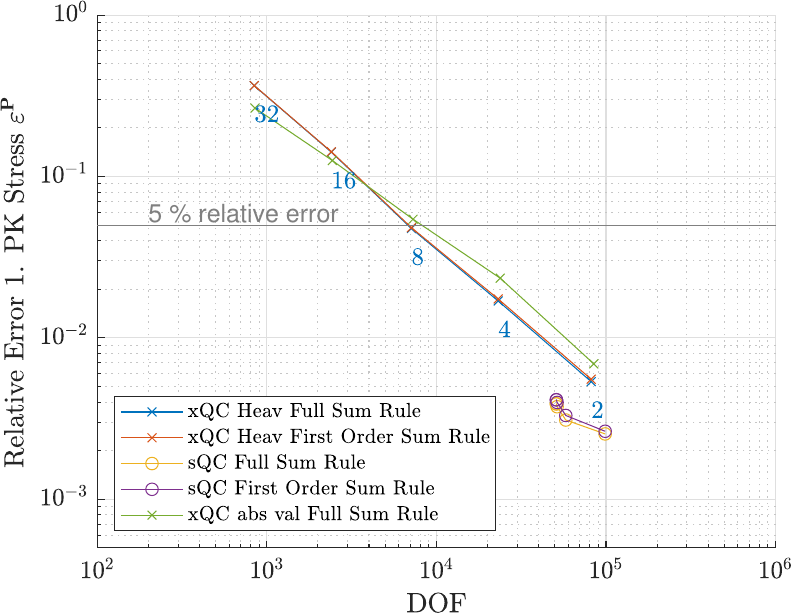}}
   \hfill
   \subfloat[Homogenized Stiffness]{\includegraphics[width=0.45\textwidth]{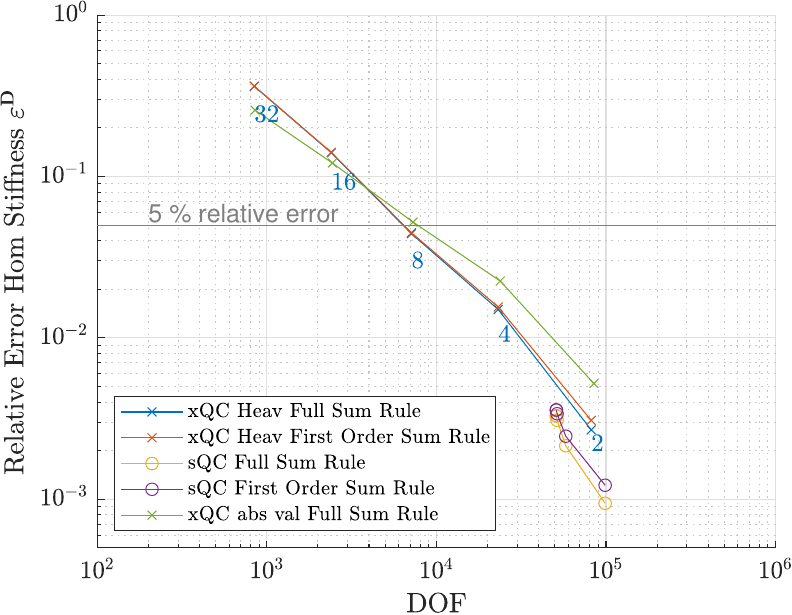}}
   \caption{The relative error of (a) the first Piola-Kirchhoff stress tensor $\mathbf{P}$ and (b) the homogenized stiffness tensor $\mathbf{D}$ of the multiple inclusion and fiber example using the extended QC with the Heaviside (Heav) and the absolute value (abs val) enrichment and the standard QC with full and first-order summation rule.}
   \label{fig:31InclConc42Fib_ErPD1212}
\end{figure}
The standard QC is predicting both tensors with an error smaller than \SI{0.4}{\%} for all triangulations (Figure \ref{fig:31InclConc42Fib_ErPD1212}). On the other hand, the extended QC leads to a range of relative error between \SI{37}{\%} and \SI{0.3}{\%}, decreasing with a smaller element size. Nevertheless, the discretization with an element size of \SI{8}{mm} is predicting the elastic energy, the homogenized stress and homogenized stiffness below the \SI{5}{\%} threshold, which is acceptable for engineering practice. Furthermore, the error of the displacement magnitude for the same numerical model is around \SI{1}{\%} by using only \SI{14}{\%} of the number of DOFs and \SI{35}{\%} of sampling interactions compared to the coarsest standard QC model. Moreover, the numerical model of the extended QC with an element size of \SI{8}{mm} uses only \SI{2.4}{\%} of DOFs and \SI{9.2}{\%} of sample interactions compared to the full model.

\section{Summary and Conclusion}
In the present study the Quasicontinuum Method was extended by enrichment strategies to account for weak discontinuities with nonconforming triangulations. This allows to investigate complex microstructures with densely packed inclusions and randomly shaped interfaces to represent them accurately while using a regular coarser mesh and reduce the computational burden. The enrichment strategies include the Heaviside, absolute value, modified absolute value and corrected XFEM. Furthermore, the first-order summation rule has been applied for the standard QC and was expanded to account for different domains inside one element cut by a material interface for the extended QC. Based on the comparative studies performed for single and multiple fiber-inclusion composites we conclude that:
\begin{enumerate}
    \item The Heaviside function has a slight advantage in accuracy of the example with a single stiff inclusion compared to the other three enrichment strategies. In the example with multiple inclusions and fibers the Heaviside enrichment leads to similar outcome as the absolute value. At the same time, the Heaviside function is the most straightforward to implement and is able to account for randomly shaped inclusions in the form of the sign function and for fibers as step function. 
    \item The first-order interaction based summation rule has been developed for the extended QC and reduces the number of sampling interactions substantially for both types of QC with less than \SI{0.1}{\%} difference compared to the full summation rule. The first-order summation rule complements the Heaviside enrichment of the extended QC, since the Heaviside function requires only the selection of one set of four interactions at one location inside the element.
    \item The extended QC with an element size of \SI{8}{mm} is predicting the elastic energy as well as the homogenized stress and stiffness tensor below the \SI{5}{\%} threshold of relative error, acceptable in many engineering applications. The extended QC model uses almost one order of magnitude fewer DOFs and one third of sampling interactions compared to the standard QC. The extended QC uses only \SI{2.4}{\%} DOFs and \SI{9.2}{\%} of sampling interactions compared to the full model.
    \item The extended QC with \SI{8}{mm} element size predicts displacements with \SI{1}{\%} relative error by using \SI{14}{\%} of DOFs and one third of sampling interactions in contrast to the standard QC. Compared to the full model only \SI{2.4}{\%} DOFs and \SI{9.2}{\%} of sampling interactions are needed.
\end{enumerate}
The extended QC developed in this manuscript can account for heterogeneous lattice networks with geometrically complex microstructures. The method can be easily combined with enrichments for strong discontinuities to predict fracture. Higher order interpolation or meshless approaches in combination with enrichment might improve the computational efficiency of the extended QC for heterogeneous lattices even further. Both topics lie within our interest and will be reported separately.

\section*{CRediT author statement}
{\bf Benjamin Werner}: Methodology, Software, Validation, Formal analysis, Investigation, Writing--Original Draft, Visualization;
{\bf Ondřej Rokoš}: Conceptualization, Methodology, Software, Formal analysis, Resources, Writing--Review \& Editing, Supervision, Funding acquisition;
{\bf Jan Zeman}: Conceptualization, Methodology, Formal analysis, Writing--Review \& Editing, Project administration, Funding acquisition.

\section*{Acknowledgement} The work of Benjamin Werner received funding from projects No.~CZ.02.2.69/0.0/0.0/ 18\_053/0016980 awarded by the Ministry of Education, Youth and Sports of the Czech Republic (from 02/2021--04/2022), No.~22-35755K awarded the Czech Science Foundation (from 01/2023) and by the CTU Global Fellowship Program (from 05/2022). The work of Ondřej Rokoš and Jan Zeman was supported by project No.~19-26143X awarded by the Czech Science Foundation.

\appendix
\section{Periodic Boundary Conditions}
\label{AppendixA}
For the implementation the periodicity conditions in Eq. (\ref{eq:period_cond}) are rewritten as
\begin{eqnarray}
 \mathbf{r}_{\mathrm{rep}}^{\mathrm{R}} = \mathbf{r}_{\mathrm{rep}}^{\mathrm{L}} + \mathbf{r}_{\mathrm{rep}}^{\mathrm{2}} -\mathbf{r}_{\mathrm{rep}}^{\mathrm{1}} \\
 \mathbf{r}_{\mathrm{rep}}^{\mathrm{T}} = \mathbf{r}_{\mathrm{rep}}^{\mathrm{B}} + \mathbf{r}_{\mathrm{rep}}^{\mathrm{4}}-\mathbf{r}_{\mathrm{rep}}^{\mathrm{1}}
\end{eqnarray}
with $\mathbf{r}_{\mathrm{rep}}^{1}$, $\mathbf{r}_{\mathrm{rep}}^{2}$ and $\mathbf{r}_{\mathrm{rep}}^{4}$ being the location of the corner repatoms and control points of the RVE (Figure \ref{fig:Schematic_PBC}b) on which the displacement is prescribed by
\begin{equation}
    \mathbf{r}_{\mathrm{rep}}^{k} = \mathbf{F} \cdot \mathbf{r}_{0 \, \mathrm{rep}}^{k} \qquad \qquad k = 1,2,4 .
\end{equation}
The first Piola-Kirchhoff stress tensor
\begin{equation}
    \mathbf{P} = \dfrac{1}{V_0} \sum_{k=1,2,4} \mathbf{f}_{\mathrm{DBC}}^k \otimes  \mathbf{r}_{0 \, \mathrm{rep}}^{k}
\end{equation}
is calculated through the initial volume $V_0$, reaction forces at the three corner nodes $\mathbf{f}_{\mathrm{DBC}}^k$ and their location in the initial configuration $\mathbf{r}_{0 \, \mathrm{rep}}^k$ of the RVE. 

The fourth order tangent stiffness tensor 
\begin{equation}
    \mathbf{D} = \dfrac{1}{V_0} \sum_i \sum_j (\mathbf{r}_{0 \, \mathrm{rep}}^i \mathbf{K}_{\mathrm{M}}^{ij} \mathbf{r}_{0 \, \mathrm{rep}}^j)^{LC}
    \label{eq:stiffness_tensor}
\end{equation}
is determined through a rearranged stiffness matrix $\mathbf{K}_{\mathrm{M}}^{ij}$ and the location of the corner repatoms $\mathbf{r}_{0 \, \mathrm{rep}}^i$ in the initial configuration. In addition, $LC$ is indicating the left conjugation and is defined as $T_{ijkl}=T_{jikl}$ for a fourth-order tensor \cite{kouznetsova20091}. The stiffness matrix $\mathbf{K}_{\mathrm{M}}^{ij}$ follows from the condensation of the system of equations 
\begin{equation}
    \begin{bmatrix}
        \mathbf{K}_{pp} & \mathbf{K}_{fp} \\
        \mathbf{K}_{pf} & \mathbf{K}_{ff}
    \end{bmatrix}
    \begin{bmatrix}
        \delta \mathbf{u}_{p}  \\
        \delta \mathbf{u}_{f}
    \end{bmatrix}
    =
    \begin{bmatrix}
        \delta \mathbf{f}_{p}  \\
        \mathbf{0}
    \end{bmatrix}
\end{equation}
with $\delta \mathbf{u}_{p}$ being the prescribed displacement increments on the three corner nodes and $\delta \mathbf{f}_{p}$ the corresponding reaction force increments. On the other hand, $\delta \mathbf{u}_{f}$ contains all remaining and therefore free DOFs and the stiffness matrix is partitioned accordingly. The unknown displacement increments $\delta \mathbf{u}_{f}$ can be solved for and substituted back leading to the expression
\begin{equation}
    \mathbf{K}_{\mathrm{M}} \delta \mathbf{u}_{p} = \delta \mathbf{f}_{p} \qquad \mathrm{with} \qquad
    \mathbf{K}_{\mathrm{M}} = \mathbf{K}_{pp} - \mathbf{K}_{pf}(\mathbf{K}_{ff})^{-1}\mathbf{K}_{fp}
\end{equation}
of the stiffness matrix. In a last step the reduced stiffness matrix $\mathbf{K}_{\mathrm{M}}$ is rewritten in a format of three $2 \times 2$ submatrices satisfying
\begin{equation}
    \sum_j \mathbf{K}_{\mathrm{M}}^{ij} \cdot \delta \mathbf{u}_{p}^j = \delta \mathbf{f}_{p}^i \qquad \mathrm{with} \qquad i,j = 1,2,4
\end{equation}
and substituted in Eq. (\ref{eq:stiffness_tensor}). The derivation of the first-order Piola-Kirchhoff stress tensor and the homogenized fourth-order stiffness tensor is described in details by Kouznetsova et al.~\cite{kouznetsova20091}.

\renewcommand*{\doi}[1]{\href{http://dx.doi.org/#1}{#1}}
\bibliographystyle{elsarticle-num-names}
\bibliography{references}

\begin{thebibliography}{55}
\expandafter\ifx\csname natexlab\endcsname\relax\def\natexlab#1{#1}\fi
\providecommand{\url}[1]{\texttt{#1}}
\providecommand{\href}[2]{#2}
\providecommand{\path}[1]{#1}
\providecommand{\DOIprefix}{doi:}
\providecommand{\ArXivprefix}{arXiv:}
\providecommand{\URLprefix}{URL: }
\providecommand{\Pubmedprefix}{pmid:}
\providecommand{\doi}[1]{\href{http://dx.doi.org/#1}{\path{#1}}}
\providecommand{\Pubmed}[1]{\href{pmid:#1}{\path{#1}}}
\providecommand{\bibinfo}[2]{#2}
\ifx\xfnm\relax \def\xfnm[#1]{\unskip,\space#1}\fi
\bibitem[{Nikolić et~al.(2018)Nikolić, Karavelić, Ibrahimbegovic, and
  Miščević}]{Nikolic2018753}
\bibinfo{author}{M.~Nikolić}, \bibinfo{author}{E.~Karavelić},
  \bibinfo{author}{A.~Ibrahimbegovic}, \bibinfo{author}{P.~Miščević},
\newblock \bibinfo{title}{{Lattice Element Models and Their Peculiarities}},
\newblock \bibinfo{journal}{Archives of Computational Methods in Engineering}
  \bibinfo{volume}{25} (\bibinfo{year}{2018}) \bibinfo{pages}{753–784}.
  \DOIprefix\doi{10.1007/s11831-017-9210-y}.
\bibitem[{Nikolić and Ibrahimbegovic(2015)}]{NIKOLIC2015209}
\bibinfo{author}{M.~Nikolić}, \bibinfo{author}{A.~Ibrahimbegovic},
\newblock \bibinfo{title}{{Rock mechanics model capable of representing initial
  heterogeneities and full set of 3D failure mechanisms}},
\newblock \bibinfo{journal}{Computer Methods in Applied Mechanics and
  Engineering} \bibinfo{volume}{290} (\bibinfo{year}{2015})
  \bibinfo{pages}{209--227}. \DOIprefix\doi{10.1016/j.cma.2015.02.024}.
\bibitem[{Bolander et~al.(2021)Bolander, Eliáš, Cusatis, and
  Nagai}]{BOLANDER2021108030}
\bibinfo{author}{J.~E. Bolander}, \bibinfo{author}{J.~Eliáš},
  \bibinfo{author}{G.~Cusatis}, \bibinfo{author}{K.~Nagai},
\newblock \bibinfo{title}{{Discrete mechanical models of concrete fracture}},
\newblock \bibinfo{journal}{Engineering Fracture Mechanics}
  \bibinfo{volume}{257} (\bibinfo{year}{2021}) \bibinfo{pages}{108030}.
  \DOIprefix\doi{10.1016/j.engfracmech.2021.108030}.
\bibitem[{Eli{\'{a}}{\v{s}} et~al.(2015)Eli{\'{a}}{\v{s}}, Vořechovsk{\'{y}},
  Sko{\v{c}}ek, and Ba{\v{z}}ant}]{Elias2015}
\bibinfo{author}{J.~Eli{\'{a}}{\v{s}}},
  \bibinfo{author}{M.~Vořechovsk{\'{y}}}, \bibinfo{author}{J.~Sko{\v{c}}ek},
  \bibinfo{author}{Z.~P. Ba{\v{z}}ant},
\newblock \bibinfo{title}{{Stochastic discrete meso-scale simulations of
  concrete fracture: Comparison to experimental data}},
\newblock \bibinfo{journal}{Engineering Fracture Mechanics}
  \bibinfo{volume}{135} (\bibinfo{year}{2015}) \bibinfo{pages}{1--16}.
  \DOIprefix\doi{10.1016/j.engfracmech.2015.01.004}.
\bibitem[{Chang et~al.(2020)Chang, Zhang, Schlangen, and
  Šavija}]{Chang20204822}
\bibinfo{author}{Z.~Chang}, \bibinfo{author}{H.~Zhang},
  \bibinfo{author}{E.~Schlangen}, \bibinfo{author}{B.~Šavija},
\newblock \bibinfo{title}{{Lattice Fracture Model for Concrete Fracture
  Revisited: Calibration and Validation}},
\newblock \bibinfo{journal}{Applied Sciences} \bibinfo{volume}{10}
  (\bibinfo{year}{2020}). \DOIprefix\doi{10.3390/app10144822}.
\bibitem[{Lilliu(2007)}]{Lilliu2007}
\bibinfo{author}{G.~Lilliu}, \bibinfo{title}{3D analysis of fracture processes
  in concrete}, Ph.D. thesis, TU Delft, \bibinfo{year}{2007}. \URLprefix
  \url{https://research.tudelft.nl/en/publications/3d-analysis-of-fracture-processes-in-concrete}.
\bibitem[{Grassl and Jir{\'{a}}sek(2010)}]{Grassl2010}
\bibinfo{author}{P.~Grassl}, \bibinfo{author}{M.~Jir{\'{a}}sek},
\newblock \bibinfo{title}{{Meso-scale approach to modelling the fracture
  process zone of concrete subjected to uniaxial tension}},
\newblock \bibinfo{journal}{International Journal of Solids and Structures}
  \bibinfo{volume}{47} (\bibinfo{year}{2010}) \bibinfo{pages}{957--968}.
  \DOIprefix\doi{10.1016/j.ijsolstr.2009.12.010}.
\bibitem[{Cusatis et~al.(2006)Cusatis, Ba{\v{z}}ant, and Cedolin}]{Cusatis2006}
\bibinfo{author}{G.~Cusatis}, \bibinfo{author}{Z.~P. Ba{\v{z}}ant},
  \bibinfo{author}{L.~Cedolin},
\newblock \bibinfo{title}{{Confinement-shear lattice CSL model for fracture
  propagation in concrete}},
\newblock \bibinfo{journal}{Computer Methods in Applied Mechanics and
  Engineering} \bibinfo{volume}{195} (\bibinfo{year}{2006})
  \bibinfo{pages}{7154--7171}. \DOIprefix\doi{10.1016/j.cma.2005.04.019}.
\bibitem[{Mayya et~al.(2016)Mayya, Praveen, Banerjee, and Rajesh}]{Mayya2016}
\bibinfo{author}{A.~Mayya}, \bibinfo{author}{P.~Praveen},
  \bibinfo{author}{A.~Banerjee}, \bibinfo{author}{R.~Rajesh},
\newblock \bibinfo{title}{{Splitting fracture in bovine bone using a
  porosity-based spring network model}},
\newblock \bibinfo{journal}{Journal of the Royal Society Interface}
  \bibinfo{volume}{13} (\bibinfo{year}{2016}).
  \DOIprefix\doi{10.1098/rsif.2016.0809}.
\bibitem[{Wilbrink et~al.(2013)Wilbrink, Beex, and Peerlings}]{Wilbrink2013}
\bibinfo{author}{D.~V. Wilbrink}, \bibinfo{author}{L.~A.~A. Beex},
  \bibinfo{author}{R.~H.~J. Peerlings},
\newblock \bibinfo{title}{{A discrete network model for bond failure and
  frictional sliding in fibrous materials}},
\newblock \bibinfo{journal}{International Journal of Solids and Structures}
  \bibinfo{volume}{50} (\bibinfo{year}{2013}) \bibinfo{pages}{1354--1363}.
  \DOIprefix\doi{10.1016/j.ijsolstr.2013.01.012}.
\bibitem[{Beex et~al.(2013)Beex, Verberne, and Peerlings}]{Beex2013}
\bibinfo{author}{L.~A.~A. Beex}, \bibinfo{author}{C.~W. Verberne},
  \bibinfo{author}{R.~H.~J. Peerlings},
\newblock \bibinfo{title}{{Experimental identification of a lattice model for
  woven fabrics: Application to electronic textile}},
\newblock \bibinfo{journal}{Composites Part A: Applied Science and
  Manufacturing} \bibinfo{volume}{48} (\bibinfo{year}{2013})
  \bibinfo{pages}{82--92}. \DOIprefix\doi{10.1016/j.compositesa.2012.12.014}.
\bibitem[{Liu et~al.(2010)Liu, Chen, and Li}]{Liu2010}
\bibinfo{author}{J.~X. Liu}, \bibinfo{author}{Z.~T. Chen},
  \bibinfo{author}{K.~C. Li},
\newblock \bibinfo{title}{{A 2-D lattice model for simulating the failure of
  paper}},
\newblock \bibinfo{journal}{Theoretical and Applied Fracture Mechanics}
  \bibinfo{volume}{54} (\bibinfo{year}{2010}) \bibinfo{pages}{1--10}.
  \DOIprefix\doi{10.1016/j.tafmec.2010.06.009}.
\bibitem[{Tadmor et~al.(1996{\natexlab{a}})Tadmor, Ortiz, and
  Phillips}]{Tadmor1996}
\bibinfo{author}{E.~B. Tadmor}, \bibinfo{author}{M.~Ortiz},
  \bibinfo{author}{R.~Phillips},
\newblock \bibinfo{title}{{Quasicontinuum analysis of defects in solids}},
\newblock \bibinfo{journal}{Philosophical Magazine A: Physics of Condensed
  Matter, Structure, Defects and Mechanical Properties} \bibinfo{volume}{73}
  (\bibinfo{year}{1996}{\natexlab{a}}) \bibinfo{pages}{1529--1563}.
  \DOIprefix\doi{10.1080/01418619608243000}.
\bibitem[{Tadmor et~al.(1996{\natexlab{b}})Tadmor, Phillips, and
  Ortiz}]{Tadmor1996a}
\bibinfo{author}{E.~B. Tadmor}, \bibinfo{author}{R.~Phillips},
  \bibinfo{author}{M.~Ortiz},
\newblock \bibinfo{title}{{Mixed atomistic and continuum models of deformation
  in solids}},
\newblock \bibinfo{journal}{Langmuir} \bibinfo{volume}{12}
  (\bibinfo{year}{1996}{\natexlab{b}}) \bibinfo{pages}{4529--4532}.
  \DOIprefix\doi{10.1021/la9508912}.
\bibitem[{Miller and Tadmor(2002)}]{Miller2002}
\bibinfo{author}{R.~E. Miller}, \bibinfo{author}{E.~B. Tadmor},
\newblock \bibinfo{title}{{The Quasicontinuum Method: Overview, applications
  and current directions}},
\newblock \bibinfo{journal}{Journal of Computer-Aided Materials Design}
  \bibinfo{volume}{9} (\bibinfo{year}{2002}) \bibinfo{pages}{203--239}.
  \DOIprefix\doi{10.1023/A:1026098010127}.
\bibitem[{Curtin and Miller(2003)}]{Curtin2003}
\bibinfo{author}{W.~A. Curtin}, \bibinfo{author}{R.~E. Miller},
\newblock \bibinfo{title}{{Atomistic/continuum coupling in computational
  materials science}},
\newblock \bibinfo{journal}{Modelling and Simulation in Materials Science and
  Engineering} \bibinfo{volume}{11} (\bibinfo{year}{2003})
  \bibinfo{pages}{R33}. \DOIprefix\doi{10.1088/0965-0393/11/3/201}.
\bibitem[{Yang et~al.(2021)Yang, Cheng, McGeough, Woldu, and Yang}]{Yang2021}
\bibinfo{author}{S.~Yang}, \bibinfo{author}{B.~Cheng}, \bibinfo{author}{J.~A.
  McGeough}, \bibinfo{author}{Y.~T. Woldu}, \bibinfo{author}{X.~Yang},
\newblock \bibinfo{title}{{Multi-scale numerical analysis and experimental
  verification for nano-cutting}},
\newblock \bibinfo{journal}{Journal of Manufacturing Processes}
  \bibinfo{volume}{71} (\bibinfo{year}{2021}) \bibinfo{pages}{260--268}.
  \DOIprefix\doi{10.1016/j.jmapro.2021.09.030}.
\bibitem[{Tran et~al.(2019)Tran, Fang, Tsai, and Chen}]{Tran2019}
\bibinfo{author}{A.~S. Tran}, \bibinfo{author}{T.~H. Fang},
  \bibinfo{author}{L.~R. Tsai}, \bibinfo{author}{C.~H. Chen},
\newblock \bibinfo{title}{{Friction and scratch characteristics of textured and
  rough surfaces using the quasi-continuum method}},
\newblock \bibinfo{journal}{Journal of Physics and Chemistry of Solids}
  \bibinfo{volume}{126} (\bibinfo{year}{2019}) \bibinfo{pages}{180--188}.
  \DOIprefix\doi{10.1016/j.jpcs.2018.11.005}.
\bibitem[{Beex et~al.(2015)Beex, Peerlings, {Van Os}, and Geers}]{Beex2015}
\bibinfo{author}{L.~A.~A. Beex}, \bibinfo{author}{R.~H.~J. Peerlings},
  \bibinfo{author}{K.~{Van Os}}, \bibinfo{author}{M.~G.~D. Geers},
\newblock \bibinfo{title}{{The mechanical reliability of an electronic textile
  investigated using the virtual-power-based quasicontinuum method}},
\newblock \bibinfo{journal}{Mechanics of Materials} \bibinfo{volume}{80}
  (\bibinfo{year}{2015}) \bibinfo{pages}{52--66}.
  \DOIprefix\doi{10.1016/j.mechmat.2014.08.001}.
\bibitem[{Roko{\v{s}} et~al.(2016)Roko{\v{s}}, Beex, Zeman, and
  Peerlings}]{Rokos2016}
\bibinfo{author}{O.~Roko{\v{s}}}, \bibinfo{author}{L.~A.~A. Beex},
  \bibinfo{author}{J.~Zeman}, \bibinfo{author}{R.~H.~J. Peerlings},
\newblock \bibinfo{title}{{A variational formulation of dissipative
  quasicontinuum methods}},
\newblock \bibinfo{journal}{International Journal of Solids and Structures}
  \bibinfo{volume}{102-103} (\bibinfo{year}{2016}) \bibinfo{pages}{214--229}.
  \DOIprefix\doi{10.1016/j.ijsolstr.2016.10.003}.
\bibitem[{Mike{\v{s}} and Jir{\'{a}}sek(2017)}]{Mikes2017}
\bibinfo{author}{K.~Mike{\v{s}}}, \bibinfo{author}{M.~Jir{\'{a}}sek},
\newblock \bibinfo{title}{{Quasicontinuum method extended to irregular
  lattices}},
\newblock \bibinfo{journal}{Computers and Structures} \bibinfo{volume}{192}
  (\bibinfo{year}{2017}) \bibinfo{pages}{50--70}.
  \DOIprefix\doi{10.1016/j.compstruc.2017.07.002}.
\bibitem[{Phlipot and Kochmann(2019)}]{Phlipot2018}
\bibinfo{author}{G.~P. Phlipot}, \bibinfo{author}{D.~M. Kochmann},
\newblock \bibinfo{title}{{A quasicontinuum theory for the nonlinear mechanical
  response of general periodic truss lattices}},
\newblock \bibinfo{journal}{Journal of the Mechanics and Physics of Solids}
  \bibinfo{volume}{124} (\bibinfo{year}{2019}) \bibinfo{pages}{758--780}.
  \DOIprefix\doi{10.1016/J.JMPS.2018.11.014}.
\bibitem[{Chen et~al.(2021)Chen, Berke, Massart, Beex, Magliulo, and
  Bordas}]{Chen2021}
\bibinfo{author}{L.~Chen}, \bibinfo{author}{P.~Z. Berke},
  \bibinfo{author}{T.~J. Massart}, \bibinfo{author}{L.~A. Beex},
  \bibinfo{author}{M.~Magliulo}, \bibinfo{author}{S.~P. Bordas},
\newblock \bibinfo{title}{{A refinement indicator for adaptive quasicontinuum
  approaches for structural lattices}},
\newblock \bibinfo{journal}{International Journal for Numerical Methods in
  Engineering} \bibinfo{volume}{122} (\bibinfo{year}{2021})
  \bibinfo{pages}{2498--2527}. \DOIprefix\doi{10.1002/nme.6629}.
\bibitem[{Chen et~al.(2020)Chen, Beex, Berke, Massart, and Bordas}]{Chen2020}
\bibinfo{author}{L.~Chen}, \bibinfo{author}{L.~A.~A. Beex},
  \bibinfo{author}{P.~Z. Berke}, \bibinfo{author}{T.~J. Massart},
  \bibinfo{author}{S.~P.~A. Bordas},
\newblock \bibinfo{title}{{Generalized quasicontinuum modeling of metallic
  lattices with geometrical and material nonlinearity and variability}},
\newblock \bibinfo{journal}{Computer Methods in Applied Mechanics and
  Engineering} \bibinfo{volume}{366} (\bibinfo{year}{2020})
  \bibinfo{pages}{112878}. \DOIprefix\doi{10.1016/j.cma.2020.112878}.
\bibitem[{Knap and Ortiz(2001)}]{Knap2001}
\bibinfo{author}{J.~Knap}, \bibinfo{author}{M.~Ortiz},
\newblock \bibinfo{title}{{An analysis of the quasicontinuum method}},
\newblock \bibinfo{journal}{Journal of the Mechanics and Physics of Solids}
  \bibinfo{volume}{49} (\bibinfo{year}{2001}) \bibinfo{pages}{1899--1923}.
  \DOIprefix\doi{10.1016/S0022-5096(01)00034-5}.
\bibitem[{Ghareeb and Elbanna(2020)}]{Ghareeb2020}
\bibinfo{author}{A.~Ghareeb}, \bibinfo{author}{A.~Elbanna},
\newblock \bibinfo{title}{{An adaptive quasicontinuum approach for modeling
  fracture in networked materials: Application to modeling of polymer
  networks}},
\newblock \bibinfo{journal}{Journal of the Mechanics and Physics of Solids}
  \bibinfo{volume}{137} (\bibinfo{year}{2020}) \bibinfo{pages}{103819}.
  \DOIprefix\doi{10.1016/j.jmps.2019.103819}.
\bibitem[{Roko{\v{s}} et~al.(2017)Roko{\v{s}}, Peerlings, Zeman, and
  Beex}]{Rokos2017}
\bibinfo{author}{O.~Roko{\v{s}}}, \bibinfo{author}{R.~H.~J. Peerlings},
  \bibinfo{author}{J.~Zeman}, \bibinfo{author}{L.~A.~A. Beex},
\newblock \bibinfo{title}{{An adaptive variational Quasicontinuum methodology
  for lattice networks with localized damage}},
\newblock \bibinfo{journal}{International Journal for Numerical Methods in
  Engineering} \bibinfo{volume}{112} (\bibinfo{year}{2017})
  \bibinfo{pages}{174--200}. \DOIprefix\doi{10.1002/nme.5518}.
\bibitem[{Shenoy et~al.(1999)Shenoy, Miller, Tadmor, Rodney, Phillips, and
  Ortiz}]{Shenoy1999}
\bibinfo{author}{V.~B. Shenoy}, \bibinfo{author}{R.~Miller},
  \bibinfo{author}{E.~B. Tadmor}, \bibinfo{author}{D.~Rodney},
  \bibinfo{author}{R.~Phillips}, \bibinfo{author}{M.~Ortiz},
\newblock \bibinfo{title}{{An adaptive finite element approach to atomic-scale
  mechanics - The quasicontinuum method}},
\newblock \bibinfo{journal}{Journal of the Mechanics and Physics of Solids}
  \bibinfo{volume}{47} (\bibinfo{year}{1999}) \bibinfo{pages}{611--642}.
  \DOIprefix\doi{10.1016/S0022-5096(98)00051-9}.
\bibitem[{Memarnahavandi et~al.(2015)Memarnahavandi, Larsson, and
  Runesson}]{Memarnahavandi2015}
\bibinfo{author}{A.~Memarnahavandi}, \bibinfo{author}{F.~Larsson},
  \bibinfo{author}{K.~Runesson},
\newblock \bibinfo{title}{{A goal-oriented adaptive procedure for the
  quasi-continuum method with cluster approximation}},
\newblock \bibinfo{journal}{Computational Mechanics} \bibinfo{volume}{55}
  (\bibinfo{year}{2015}) \bibinfo{pages}{617--642}.
  \DOIprefix\doi{10.1007/s00466-015-1127-4}.
\bibitem[{Mo{\"{e}}s et~al.(2003)Mo{\"{e}}s, Cloirec, Cartraud, and
  Remacle}]{Moes2003}
\bibinfo{author}{N.~Mo{\"{e}}s}, \bibinfo{author}{M.~Cloirec},
  \bibinfo{author}{P.~Cartraud}, \bibinfo{author}{J.~F. Remacle},
\newblock \bibinfo{title}{{A computational approach to handle complex
  microstructure geometries}},
\newblock \bibinfo{journal}{Computer Methods in Applied Mechanics and
  Engineering} \bibinfo{volume}{192} (\bibinfo{year}{2003})
  \bibinfo{pages}{3163--3177}. \DOIprefix\doi{10.1016/S0045-7825(03)00346-3}.
\bibitem[{Fries and Belytschko(2010)}]{Fries2010}
\bibinfo{author}{T.-P. Fries}, \bibinfo{author}{T.~Belytschko},
\newblock \bibinfo{title}{{The extended/generalized finite element method: An
  overview of the method and its applications}},
\newblock \bibinfo{journal}{International Journal for Numerical Methods in
  Engineering} \bibinfo{volume}{84} (\bibinfo{year}{2010})
  \bibinfo{pages}{253--304}. \DOIprefix\doi{10.1002/nme.2914}.
\bibitem[{Cheng and Fries(2010)}]{Cheng2010}
\bibinfo{author}{K.~W. Cheng}, \bibinfo{author}{T.-P. Fries},
\newblock \bibinfo{title}{{Higher-order XFEM for curved strong and weak
  discontinuities}},
\newblock \bibinfo{journal}{International Journal for Numerical Methods in
  Engineering} \bibinfo{volume}{82} (\bibinfo{year}{2010})
  \bibinfo{pages}{564--590}. \DOIprefix\doi{10.1002/nme.2768}.
\bibitem[{Li et~al.(2018)Li, Li, and Yuan}]{Li2018}
\bibinfo{author}{H.~Li}, \bibinfo{author}{J.~Li}, \bibinfo{author}{H.~Yuan},
\newblock \bibinfo{title}{{A review of the extended finite element method on
  macrocrack and microcrack growth simulations}},
\newblock \bibinfo{journal}{Theoretical and Applied Fracture Mechanics}
  \bibinfo{volume}{97} (\bibinfo{year}{2018}) \bibinfo{pages}{236--249}.
  \DOIprefix\doi{10.1016/j.tafmec.2018.08.008}.
\bibitem[{Zhu(2012)}]{Zhu2012186}
\bibinfo{author}{Q.-Z. Zhu},
\newblock \bibinfo{title}{{On enrichment functions in the extended finite
  element method}},
\newblock \bibinfo{journal}{International Journal for Numerical Methods in
  Engineering} \bibinfo{volume}{91} (\bibinfo{year}{2012})
  \bibinfo{pages}{186--217}. \DOIprefix\doi{10.1002/nme.4272}.
\bibitem[{Wu and Li(2015)}]{WU201577}
\bibinfo{author}{J.-Y. Wu}, \bibinfo{author}{F.-B. Li},
\newblock \bibinfo{title}{{An improved stable XFEM (Is-XFEM) with a novel
  enrichment function for the computational modeling of cohesive cracks}},
\newblock \bibinfo{journal}{Computer Methods in Applied Mechanics and
  Engineering} \bibinfo{volume}{295} (\bibinfo{year}{2015})
  \bibinfo{pages}{77--107}. \DOIprefix\doi{10.1016/j.cma.2015.06.018}.
\bibitem[{Gupta et~al.(2013)Gupta, Duarte, Babuška, and
  Banerjee}]{GUPTA201323}
\bibinfo{author}{V.~Gupta}, \bibinfo{author}{C.~Duarte},
  \bibinfo{author}{I.~Babuška}, \bibinfo{author}{U.~Banerjee},
\newblock \bibinfo{title}{{A stable and optimally convergent generalized FEM
  (SGFEM) for linear elastic fracture mechanics}},
\newblock \bibinfo{journal}{Computer Methods in Applied Mechanics and
  Engineering} \bibinfo{volume}{266} (\bibinfo{year}{2013})
  \bibinfo{pages}{23--39}. \DOIprefix\doi{10.1016/j.cma.2013.07.010}.
\bibitem[{Roko{\v{s}} et~al.(2017)Roko{\v{s}}, Peerlings, and
  Zeman}]{Rokos2017x}
\bibinfo{author}{O.~Roko{\v{s}}}, \bibinfo{author}{R.~H.~J. Peerlings},
  \bibinfo{author}{J.~Zeman},
\newblock \bibinfo{title}{{eXtended variational quasicontinuum methodology for
  lattice networks with damage and crack propagation}},
\newblock \bibinfo{journal}{Computer Methods in Applied Mechanics and
  Engineering} \bibinfo{volume}{320} (\bibinfo{year}{2017})
  \bibinfo{pages}{769--792}. \DOIprefix\doi{10.1016/j.cma.2017.03.042}.
\bibitem[{Eidel and Stukowski(2009)}]{Eidel2009}
\bibinfo{author}{B.~Eidel}, \bibinfo{author}{A.~Stukowski},
\newblock \bibinfo{title}{{A variational formulation of the quasicontinuum
  method based on energy sampling in clusters}},
\newblock \bibinfo{journal}{Journal of the Mechanics and Physics of Solids}
  \bibinfo{volume}{57} (\bibinfo{year}{2009}) \bibinfo{pages}{87--108}.
  \DOIprefix\doi{10.1016/j.jmps.2008.09.017}.
\bibitem[{Yang et~al.(2013)Yang, Biyikli, and To}]{Yang2013}
\bibinfo{author}{Q.~Yang}, \bibinfo{author}{E.~Biyikli}, \bibinfo{author}{A.~C.
  To},
\newblock \bibinfo{title}{{Multiresolution molecular mechanics: Statics}},
\newblock \bibinfo{journal}{Computer Methods in Applied Mechanics and
  Engineering} \bibinfo{volume}{258} (\bibinfo{year}{2013})
  \bibinfo{pages}{26--38}. \DOIprefix\doi{10.1016/j.cma.2013.01.014}.
\bibitem[{Beex et~al.(2014)Beex, Peerlings, and Geers}]{Beex2014}
\bibinfo{author}{L.~A.~A. Beex}, \bibinfo{author}{R.~H.~J. Peerlings},
  \bibinfo{author}{M.~G.~D. Geers},
\newblock \bibinfo{title}{{Central summation in the quasicontinuum method}},
\newblock \bibinfo{journal}{Journal of the Mechanics and Physics of Solids}
  \bibinfo{volume}{70} (\bibinfo{year}{2014}) \bibinfo{pages}{242--261}.
  \DOIprefix\doi{10.1016/j.jmps.2014.05.019}.
\bibitem[{Beex et~al.(2011)Beex, Peerlings, and Geers}]{Beex2011}
\bibinfo{author}{L.~A.~A. Beex}, \bibinfo{author}{R.~H.~J. Peerlings},
  \bibinfo{author}{M.~G.~D. Geers},
\newblock \bibinfo{title}{{A quasicontinuum methodology for multiscale analyses
  of discrete microstructural models}},
\newblock \bibinfo{journal}{International Journal for Numerical Methods in
  Engineering}  (\bibinfo{year}{2011}) \bibinfo{pages}{701--718}.
  \DOIprefix\doi{10.1002/nme.3134}.
\bibitem[{Amelang et~al.(2015)Amelang, Venturini, and Kochmann}]{Amelang2015}
\bibinfo{author}{J.~S. Amelang}, \bibinfo{author}{G.~N. Venturini},
  \bibinfo{author}{D.~M. Kochmann},
\newblock \bibinfo{title}{{Summation rules for a fully nonlocal energy-based
  quasicontinuum method}},
\newblock \bibinfo{journal}{Journal of the Mechanics and Physics of Solids}
  \bibinfo{volume}{82} (\bibinfo{year}{2015}) \bibinfo{pages}{378--413}.
  \DOIprefix\doi{10.1016/j.jmps.2015.03.007}.
\bibitem[{Amelang and Kochmann(2015)}]{Amelang2015a}
\bibinfo{author}{J.~S. Amelang}, \bibinfo{author}{D.~M. Kochmann},
\newblock \bibinfo{title}{{Surface effects in nanoscale structures investigated
  by a fully-nonlocal energy-based quasicontinuum method}},
\newblock \bibinfo{journal}{Mechanics of Materials} \bibinfo{volume}{90}
  (\bibinfo{year}{2015}) \bibinfo{pages}{166--184}.
  \DOIprefix\doi{10.1016/j.mechmat.2015.04.004}.
\bibitem[{Moës et~al.(1999)Moës, Dolbow, and Belytschko}]{Moes1999}
\bibinfo{author}{N.~Moës}, \bibinfo{author}{J.~Dolbow},
  \bibinfo{author}{T.~Belytschko},
\newblock \bibinfo{title}{{A finite element method for crack growth without
  remeshing}},
\newblock \bibinfo{journal}{International Journal for Numerical Methods in
  Engineering} \bibinfo{volume}{46} (\bibinfo{year}{1999})
  \bibinfo{pages}{131--150}.
  \DOIprefix\doi{10.1002/(SICI)1097-0207(19990910)46:1<131::AID-NME726>3.0.CO;2-J}.
\bibitem[{Krongauz and Belytschko(1998)}]{Krongauz1998}
\bibinfo{author}{Y.~Krongauz}, \bibinfo{author}{T.~Belytschko},
\newblock \bibinfo{title}{{EFG approximation with discontinuous derivatives}},
\newblock \bibinfo{journal}{International Journal for Numerical Methods in
  Engineering} \bibinfo{volume}{41} (\bibinfo{year}{1998})
  \bibinfo{pages}{1215--1233}.
  \DOIprefix\doi{10.1002/(SICI)1097-0207(19980415)41:7<1215::AID-NME330>3.0.CO;2-\%23}.
\bibitem[{Fries(2008)}]{Fries2008}
\bibinfo{author}{T.-P. Fries},
\newblock \bibinfo{title}{{A corrected XFEM approximation without problems in
  blending elements}},
\newblock \bibinfo{journal}{International Journal for Numerical Methods in
  Engineering} \bibinfo{volume}{75} (\bibinfo{year}{2008})
  \bibinfo{pages}{503--532}. \DOIprefix\doi{10.1002/nme.2259}.
\bibitem[{Lang et~al.(2014)Lang, Makhija, Doostan, and Maute}]{Lang20141357}
\bibinfo{author}{C.~Lang}, \bibinfo{author}{D.~Makhija},
  \bibinfo{author}{A.~Doostan}, \bibinfo{author}{K.~Maute},
\newblock \bibinfo{title}{{A simple and efficient preconditioning scheme for
  heaviside enriched XFEM}},
\newblock \bibinfo{journal}{Computational Mechanics} \bibinfo{volume}{54}
  (\bibinfo{year}{2014}) \bibinfo{pages}{1357 – 1374}.
  \DOIprefix\doi{10.1007/s00466-014-1063-8}, \bibinfo{note}{cited by: 72; All
  Open Access, Green Open Access}.
\bibitem[{Kramer et~al.(2013)Kramer, Bochev, Siefert, and Voth}]{KRAMER201370}
\bibinfo{author}{R.~Kramer}, \bibinfo{author}{P.~Bochev},
  \bibinfo{author}{C.~Siefert}, \bibinfo{author}{T.~Voth},
\newblock \bibinfo{title}{{An extended finite element method with algebraic
  constraints (XFEM-AC) for problems with weak discontinuities}},
\newblock \bibinfo{journal}{Computer Methods in Applied Mechanics and
  Engineering} \bibinfo{volume}{266} (\bibinfo{year}{2013})
  \bibinfo{pages}{70--80}. \DOIprefix\doi{10.1016/j.cma.2013.07.013}.
\bibitem[{Radtke et~al.(2010{\natexlab{a}})Radtke, Simone, and
  Sluys}]{Radtke2010}
\bibinfo{author}{F.~K.~F. Radtke}, \bibinfo{author}{A.~Simone},
  \bibinfo{author}{L.~J. Sluys},
\newblock \bibinfo{title}{{A partition of unity finite element method for
  obtaining elastic properties of continua with embedded thin fibres}},
\newblock \bibinfo{journal}{International Journal for Numerical Methods in
  Engineering} \bibinfo{volume}{84} (\bibinfo{year}{2010}{\natexlab{a}})
  \bibinfo{pages}{708--732}. \DOIprefix\doi{10.1002/nme.2916}.
\bibitem[{Radtke et~al.(2010{\natexlab{b}})Radtke, Simone, and
  Sluys}]{Radtke2010b}
\bibinfo{author}{F.~K.~F. Radtke}, \bibinfo{author}{A.~Simone},
  \bibinfo{author}{L.~J. Sluys},
\newblock \bibinfo{title}{{A computational model for failure analysis of fibre
  reinforced concrete with discrete treatment of fibres}},
\newblock \bibinfo{journal}{Engineering Fracture Mechanics}
  \bibinfo{volume}{77} (\bibinfo{year}{2010}{\natexlab{b}})
  \bibinfo{pages}{597--620}. \DOIprefix\doi{10.1016/j.engfracmech.2009.11.014}.
\bibitem[{Radtke et~al.(2011)Radtke, Simone, and Sluys}]{Radtke2011}
\bibinfo{author}{F.~K.~F. Radtke}, \bibinfo{author}{A.~Simone},
  \bibinfo{author}{L.~J. Sluys},
\newblock \bibinfo{title}{{A partition of unity finite element method for
  simulating non-linear debonding and matrix failure in thin fibre
  composites}},
\newblock \bibinfo{journal}{International Journal for Numerical Methods in
  Engineering} \bibinfo{volume}{86} (\bibinfo{year}{2011})
  \bibinfo{pages}{453--476}. \DOIprefix\doi{10.1002/nme.3056}.
\bibitem[{Pike and Oskay(2015)}]{Pike2015}
\bibinfo{author}{M.~G. Pike}, \bibinfo{author}{C.~Oskay},
\newblock \bibinfo{title}{{Modeling Random Short Nanofiber- and
  Microfiber-Reinforced Composites Using the Extended Finite-Element Method}},
\newblock \bibinfo{journal}{Journal of Nanomechanics and Micromechanics}
  \bibinfo{volume}{5} (\bibinfo{year}{2015}) \bibinfo{pages}{1--11}.
  \DOIprefix\doi{10.1061/(asce)nm.2153-5477.0000092}.
\bibitem[{Tran et~al.(2011)Tran, Yvonnet, He, Toulemonde, and
  Sanahuja}]{Tran20111436}
\bibinfo{author}{A.~B. Tran}, \bibinfo{author}{J.~Yvonnet},
  \bibinfo{author}{Q.-C. He}, \bibinfo{author}{C.~Toulemonde},
  \bibinfo{author}{J.~Sanahuja},
\newblock \bibinfo{title}{{A multiple level set approach to prevent numerical
  artefacts incomplex microstructures with nearby inclusions within XFEM}},
\newblock \bibinfo{journal}{International Journal for Numerical Methods in
  Engineering} \bibinfo{volume}{85} (\bibinfo{year}{2011})
  \bibinfo{pages}{1436--1459}. \DOIprefix\doi{10.1002/nme.3025}.
\bibitem[{Hiriyur et~al.(2011)Hiriyur, Waisman, and Deodatis}]{Hiriyur2011257}
\bibinfo{author}{B.~Hiriyur}, \bibinfo{author}{H.~Waisman},
  \bibinfo{author}{G.~Deodatis},
\newblock \bibinfo{title}{{Uncertainty quantification in homogenization of
  heterogeneous microstructures modeled by XFEM}},
\newblock \bibinfo{journal}{International Journal for Numerical Methods in
  Engineering} \bibinfo{volume}{88} (\bibinfo{year}{2011})
  \bibinfo{pages}{257--278}. \DOIprefix\doi{10.1002/nme.3174}.
\bibitem[{Kouznetsova et~al.(2009)Kouznetsova, Geers, and
  Brekelmans}]{kouznetsova20091}
\bibinfo{author}{V.~G. Kouznetsova}, \bibinfo{author}{M.~G.~D. Geers},
  \bibinfo{author}{W.~A.~M. Brekelmans}, \bibinfo{title}{Computational
  homogenisation for non-linear heterogeneous solids}, \bibinfo{year}{2009}.
  \DOIprefix\doi{10.1142/9781848163089\_0001}.

\end{thebibliography}

\end{document}